\documentclass[conference]{IEEEtran}
\usepackage{multirow}
\usepackage{url}
\usepackage{amsmath}
\usepackage{array}
\usepackage{amssymb}
\usepackage{amsfonts}
\newtheorem{proposition}{Proposition}

\newtheorem{definition}{Definition}
\newtheorem{theorem}{Theorem}

\title{Fast-Group-Decodable STBCs via Codes over GF(4)}
\author{
\authorblockN{N.~Lakshmi Prasad and B.~Sundar Rajan}
\authorblockA{Dept. of ECE, IISc, Bangalore 560012, India\\
Email: \{nlp,bsrajan\}@ece.iisc.ernet.in\\}
}
\date{\today}
\begin{document}
\maketitle
\thispagestyle{empty}

\begin{abstract}
In this paper we construct low decoding complexity STBCs by using the Pauli matrices as linear dispersion matrices. In this case the Hurwitz-Radon orthogonality condition is shown to be easily checked by transferring the problem to $\mathbb{F}_4$ domain. The problem of constructing low decoding complexity STBCs is shown to be equivalent to finding certain codes over $\mathbb{F}_4$. It is shown that almost all known low complexity STBCs can be obtained by this approach. New codes are given that have the least known decoding complexity in particular ranges of rate.
\end{abstract}
\section{Introduction}
Consider an $N$ transmit antenna, $N_r$ receive antenna quasi-static Rayleigh flat fading MIMO channel given by
\begin{equation} \label{MIMOChannel}
Y = \mathfrak{X}H + W,
\end{equation}
where $H$ is the  $N \times N_r$ channel matrix, $\mathfrak{X}$ is the $T \times N$ matrix of transmitted signal, $W$ is the $T \times N_r$ additive noise matrix and $Y$ is the $T \times N_r$ matrix of received signal, all matrices being over the complex field $\mathbb{C}.$ Throughout this paper, we consider only the case $T=N.$

An $N \times N$ Space-Time Block Code (STBC) $\mathcal{C}$ is a finite subset of $\mathbb{C}^{N \times N}$. An $N \times N$ Linear Space-Time Design (LSTD) \cite{HaH} or simply a design ${\bf X}$ in $K$ real variables \mbox{$x_1,\dots,x_K$} is a matrix \begin{equation}\label{lstd} \sum_{i=1}^{K}{x_iA_i} \end{equation} where \mbox{$A_i \in \mathbb{C}^{N \times N}$}, $i=1,\dots,K$ and the set \mbox{$\{A_1,\dots,A_K\}$} is linearly independent over the field of real numbers $\mathbb{R}$. Throughout the paper designs are denoted using upper case bold font letters. An $N \times N$ design ${\bf X}$ in $K$ real variables is said to have a rate of \mbox{$R=\frac{K}{2N}$} complex symbols per channel use (cspcu). The matrices $A_i$ are known as linear dispersion or weight matrices. An STBC can be obtained from a design ${\bf X}$ by making \mbox{$x_1,\dots,x_K$} take values from a finite set \mbox{$\mathcal{A} \subset \mathbb{R}^K$}. The set $\mathcal{A}$ is called the signal set. Denote the STBC obtained this way by $\mathcal{C}({\bf X},\mathcal{A})$. Thus we have
\begin{equation} \label{stbc}
\mathcal{C}({\bf X},\mathcal{A}) = \{ \sum_{l=1}^{K}{a_lA_l} | [a_1,\dots,a_K]^T \in \mathcal{A} \}
\end{equation}

One of the important aspects in the design of space-time block codes (STBC) is the ML decoding complexity. Space-Time block codes based on orthogonal designs were proposed in \cite{TJC},\cite{Lia},\cite{TiH}. Clifford Algebras were proposed as a means to design square orthogonal designs in \cite{TiH}. It is known that \cite{YGT2},\cite{WWX},\cite{KhR} orthogonal designs offer single real symbol decodability and full diversity. However the rates offered by these designs is less than $1$ complex symbol per channel use when the number of transmit antennas is more than two \cite{TJC},\cite{Lia},\cite{TiH}. Single complex symbol decodable or double real symbol decodable rate $1$ STBCs were given in \cite{YGT2},\cite{WWX},\cite{KhR},\cite{KaR2}. 

In \cite{DYTGT},\cite{KaR2},\cite{KaR1} the framework for multigroup decodable STBCs was given. In \eqref{lstd}, if the symbols \mbox{$x_1,\dots,x_K$} can be divided into groups such that each group of symbols can be ML decoded independently of other groups, then the decoding complexity is greatly reduced. A necessary condition for $x_i$ and $x_j$ to be decoded independently of each other is that their weight matrices $A_i$ and $A_j$ must be Hurwitz-Radon orthogonal, i.e., they must satisfy
\begin{equation} \label{Hurwitz-Radon-orthogonality}
A_i^HA_j + A_j^HA_i = {\bf 0}
\end{equation}
Constructing low decoding complexity STBCs requires one to find weight matrices satisfying the above equation. In \cite{DYTGT} $4$-group decodable rate $1$ codes for arbitrary antennas was given. Coordinate Interleaved Orthogonal Designs or CIODs \cite{KhR} provide single complex symbol decodable designs for arbitrary number of transmit antennas. However their rate decreases rapidly with increasing number of antennas. \cite{KaR2} gave a general algebraic structure of the weight matrices of $g$-group decodable codes using Clifford Algebras, where different groups can have different number of information symbols to be decoded together. \cite{KaR1} gives $g$-group decodable designs, called Clifford Unitary Weight Designs (CUWDs), by manipulating the matrices obtained through representation of Clifford Algebras. In \cite{RaR} an algebraic framework based on Extended Clifford Algebras was proposed to study CUW STBCs and using this framework, the optimal tradeoff between rate and ML decoding complexity of CUW STBCs was obtained for few specific cases. Codes meeting this tradeoff were also provided. 

In~\cite{BHV} fast-decodable(FD) codes were introduced. These codes were not multigroup decodable but they had reduced sphere-decoding complexity. Later in~\cite{SrR1},~\cite{SiB} it was shown that the Golden Code~\cite{BRV} is fast-decodable and hence has lower ML decoding complexity than previously thought of. In~\cite{SrR1} rate $2$ codes for $2$, $4$ antennas with the largest known coding gain were given. These codes too were fast-decodable. Recently fast-decodable codes for number of antennas $N=2,4,6,8$ were given in~\cite{SBM} . These codes, known as EAST codes, combine a modified version of Perfect codes~\cite{ORBV} with Alamouti embedding. Their rates are limited to $1\leq R \leq N/2$ complex symbols per channel use, but they posses the non-vanishing determinant property. In~\cite{RGYS} a new class of codes called fast-group-decodable(FGD) were introduced. These codes combine the low decoding complexity properties of multigroup and FD codes. 


Let $ X=\begin{pmatrix}
0 & 1 \\
1 & 0
\end{pmatrix}$ 
and $ Z=\begin{pmatrix}
1 & 0 \\
0 & -1
\end{pmatrix}$. Note that both $X$ and $Z$ are Hermitian and unitary. The four matrices $I_2,X,Z$ and $iXZ$ are known as the Pauli matrices. They form a $\mathbb{C}$-linear basis of $\mathbb{C}^{2 \times 2}$. The finite group $G_m$ generated by the $m^{th}$ order tensor products of the Pauli matrices is called the Pauli group. It consists of all possible $m$ fold tensor products of the Pauli matrices together with multiplicative factors $\pm 1,\pm i,$ i.e.,
\begin{equation*}
G_m = \lbrace ~ i^{\mu}B_1 \otimes \cdots \otimes B_{m} \vert \mu \in \mathbb{Z}_4 \textrm{ and } B_k \in \{I_2,X,Z,iXZ \} ~ \rbrace .
\end{equation*}
The following  subest of $G_m,$ 
\begin{equation*}
\Lambda_m = \lbrace ~ i^{\lambda}B_1 \otimes \cdots \otimes B_{m} \vert \lambda \in \mathbb{Z}_2 \textrm{ and } B_k \in \{I_2,iX,iZ,ZX \} ~ \rbrace
\end{equation*}
is a basis for $\mathbb{C}^{2^m \times 2^m}$ as a vector space over $\mathbb{R}$.

We relate the set $\Lambda_m$ to a subset of $\mathbb{F}^{m+1}_4,$ where $\mathbb{F}_4$ is the finite field with $4$ elements \mbox{$\lbrace 0,1,\omega,\omega^2 \rbrace$} satisfying the relations \mbox{$1 + \omega = \omega^2$,} by defining the map
\begin{equation*}
\psi : \{I_2,iX,iZ,ZX\} \to \mathbb{F}_4.
\end{equation*}
that sends
\begin{align}
& I_2 \to 0 \textrm{, } iX \to 1 \textrm{, } iZ \to \omega \textrm{, } ZX \to \omega^2.
\end{align}
Note that every element \mbox{$t \in \Lambda_m$} can be uniquely written as
\begin{equation}
t = i^\lambda\psi^{-1}(\xi_1) \otimes \cdots \otimes \psi^{-1}(\xi_m) \textrm{ for } \lambda \in \mathbb{Z}_2 \textrm{ and } \xi_k \in \mathbb{F}_4.
\end{equation}
In fact, the map $\varphi : \Lambda_m \to \mathbb{F}_2 \oplus \mathbb{F}_4^m$ that sends
        \begin{equation*}
        i^{\lambda}B_1 \otimes \cdots \otimes B_{m} \to [\lambda,\psi(B_1),\dots,\psi(B_m)]
        \end{equation*}
is a one to one correspondence between $\Lambda_m$ and $\mathbb{F}_2 \oplus \mathbb{F}_4^m$. The Hamming weight ${\sf wt}([\lambda,\xi_1,\dots,\xi_m])$ of a vector \mbox{$[\lambda,\xi_1,\dots,\xi_m] \in \mathbb{F}_2 \oplus \mathbb{F}_4^m$} is defined as         \begin{equation*}         {\sf wt}([\lambda,\xi_1,\dots,\xi_m]) = {\sf \bf 1}\{ \lambda \neq 0 \} + \sum_{i=1}^{m}{{\sf \bf 1}\{ \xi_i \neq 0 \}}.         \end{equation*}

In this paper we construct low decoding complexity STBCs from designs that use elements from the Pauli group as linear dispersion matrices. This problem is converted to one of finding certain codes over $\mathbb{F}_4,$ which are subsets of $\mathbb{F}_2 \oplus \mathbb{F}_4^m,$ in which checking the conditions for low decodability becomes simpler. The contributions of this paper are as follows.
\begin{itemize}

\item It is shown that when designs are constructed by using elements of Pauli group as weight matrices, the Hurwitz-Radon orthogonality condition~\eqref{Hurwitz-Radon-orthogonality} can be easily checked by transferring the problem to the corresponding $\mathbb{F}_4$-domain. This facilitates both the description and the construction of low decoding complexity codes in the $\mathbb{F}_4$-domain. 

\item A new class of full-diversity fast-group-decodable and fast-decodable codes is reported. The codes of this class have the lowest decoding complexity among all known codes for rates \mbox{$1 < R < 3/2$} and \mbox{$2^{m-2}+\frac{1}{2^m}<R\leq2^m$} for $2^m$ antennas. Particular examples of codes from this class are
	\begin{itemize}
	\item Code for $N=4$ transmit antennas, rate $R=5/4$ with a complexity of the order of $M^2$ for arbitrary constellations and $M^{1.5}$ for constellations carefully chosen to reduce decoding complexity. This is in comparison with the code given in~\cite{YGT1} and~\cite{SrR2} with complexity $M^{2.5}$ for arbitrary constellations.
	\item Code for $N=4$ and $R=17/8$ with a complexity of the order of $M^{5.5}$ and $M^5$ for arbitrary and carefully chosen constellations respectively. This is in comparison with the code given in~\cite{RGYS} with a complexity of $M^6$. It is also shown that the STBC given in~\cite{RGYS} is a specific case of STBCs obtainable from codes over $\mathbb{F}_4$.
\end{itemize}
	
\item We construct STBCs from codes over $\mathbb{F}_4$ that meet the rate-ML decoding complexity tradeoff of a particular class of CUWDs obtainable from Extended Clifford Algebras.

\item It is  shown that almost all the low decoding complexity STBCs available in the literature can be obtained via codes over $\mathbb{F}_4$.

\item We show that full-diversity STBCs with a decoding complexity of $M^{2^{m-1}R-0.5}$ can be obtained by using regular PAM constellation for a subset of real symbols in the design given in~\cite{SrR2}. For rates $3/2 \leq R \leq 2^{m-2}+\frac{1}{2^m}$ these codes have the least known decoding complexity for $2^m$ antennas.

\item STBCs obtainable by our approach are shown to include as special cases, the $R=2$ codes, proposed in~\cite{SrR1}, for $2$ and $4$ antennas having the largest known coding gain and the fast-decodable rate $2$ code for $2$ transmit antennas proposed in~\cite{HTW},~\cite{PGA}. 

\item We show that if a design has only full-rank linear dispersion matrices, then a full diversity STBC can always be constructed from it by encoding each real symbol independently.

\end{itemize}

The contents of this paper are organized as follows. In Section~\ref{BackgroundSection} preliminary results and the necessary background and notations are established. Results regarding full diversity are given in Section~\ref{FullDiversitySection}. The focus of the paper then shifts to designing low complexity designs rather than specific STBCs. Clifford Algebras and Pauli matrices are introduced in Section~\ref{CliffordAlgebrasSection} and the problem of designing low complexity designs from $G_m$ is converted to a problem in \mbox{$\mathbb{F}_4^{m+1}$}. Examples of STBCs in the literature that can be obtained via codes over $\mathbb{F}_4$ are given in this section. In Section~\ref{MultigroupSection}, it is shown that almost all known multi group decodable codes are obtainable via codes over $\mathbb{F}_4.$ Also, a new class of  multi-group decodable codes is presented. In Section~\ref{FDSection} new fast-group-decodable and fast-decodable codes are presented. Concluding remarks are made in Section~\ref{FutureWorkSection}.

\noindent
{\sf \bf Notation:} For a complex matrix $A$ the transpose, the conjugate and the conjugate-transpose are denoted by $A^T$,$\bar{A}$ and $A^H$ respectively. $||A||^2_F$ is the Frobenius norm of the matrix $A$. $A \otimes B$ is the Kronecker product of matrices $A$ and $B$. $I_n$ is the $n \times n$ identity matrix and {\bf 0 }is the all zero matrix of appropriate dimension. The empty set is denoted by $\phi$. Cardinality of a set $\Gamma$ is denoted by $\vert\Gamma\vert$ and $i = \sqrt{-1}$. ~ ${\sf \bf 1}\{\cdots\}$ is the indicator function. For a square matrix $A$, $det(A)$ is the determinant of $A$. For a positive integer $n$, $\mathbb{Z}_n$ is the set \mbox{$\{0,1,\dots,n-1\}$}. For a complex matrix $A$, $A_{Re}$ and $A_{Im}$ denote its real and imaginary parts respectively. $vec(A)$ is the vectorization of the matrix $A$.


\section{Preliminaries} \label{BackgroundSection}

In this section,  multigroup decodability, fast-decodability \cite{BHV} and fast-group-decodability \cite{RGYS} of STBCs and designs are formally defined with the help of the notion of group-decodability. 

Let $N_K$ denote the set \mbox{$\{1,\dots,K\}$}. For any $K$-tuple $x=(x_1,\dots,x_K)$ and non-empty set \mbox{$\Gamma \subseteq N_K$} define 
\begin{equation*}
x_{\Gamma}=(x_{i_1},x_{i_2},\dots,x_{i_{|\Gamma|}})
\end{equation*}
wherein \mbox{$\Gamma=\{i_1,\dots,i_{|\Gamma|}\}$}. The idea of encoding complexity was first introduced in~\cite{RaR}. We now define $g$-group encodability of an STBC $\mathcal{C}({\bf X},\mathcal{A})$.
\begin{definition}
Let $g$ be any positive integer. An STBC $\mathcal{C}({\bf X},\mathcal{A})$ obtained from a design ${\bf X}$ and a signal set $\mathcal{A}$ is said to be $g$-group encodable if there exists a partition of $N_K$ into non-empty subsets \mbox{$\Gamma_1,\dots,\Gamma_g$} and there exist finite subsets \mbox{$\mathcal{A}_{i} \subset \mathbb{R}^{|\Gamma_i|}$}, $i=1,\dots,g$ such that
	\begin{equation*}
	\mathcal{C}({\bf X},\mathcal{A}) = \{ \sum_{l=1}^{K}{a_lA_l} | a_{\Gamma_i} \in \mathcal{A}_{i} \textrm{ for } i=1,\dots,g\}
	\end{equation*}
\end{definition}

In short, for a $g$-group encodable STBC the tuples \mbox{$x_{\Gamma_1},\dots,x_{\Gamma_g}$} are assigned values independently of each other during encoding. If for each $i$, $|\Gamma_i| = 1$, we say that the STBC $\mathcal{C}({\bf X},\mathcal{A})$ is single real symbol encodable. For a design ${\bf X}$ in $K$ real variables and a non-empty subset $\Gamma \subseteq N_K$ define the design ${\bf X}_{\Gamma}$ as follows
\begin{equation*}
{\bf X}_{\Gamma} = \sum_{i \in \Gamma}{x_iA_i}
\end{equation*}

We now define the $g$-group maximum-likelihood (ML) decodability of an STBC $\mathcal{C}({\bf X},\mathcal{A})$.
\begin{definition}
An STBC $\mathcal{C}({\bf X},\mathcal{A})$ is said to be $g$-group ML decodable if there exists a partition of $N_K$ into $g$ non-empty subsets \mbox{$\Gamma_1,\dots,\Gamma_g$} and if there exist finite subsets \mbox{$\mathcal{A}_{i} \subset \mathbb{R}^{|\Gamma_i|}$}, $i=1,\dots,g$ such that the ML decoder, \mbox{$\textrm{arg min}_{C \in \mathcal{C}({\bf X},\mathcal{A})}||Y-CH||^2_F$}, decomposes as 
	\begin{equation*}
	\sum_{i=1}^{g} \textrm{arg min}_{C_i \in \mathcal{C}({\bf X_{\Gamma_i}},\mathcal{A}_i)}||Y-C_iH||^2_F
	\end{equation*}
\end{definition}

Such a decomposition reduces the decoding complexity from $\prod_{i=1}^{g}{|\mathcal{A}_i|}$ computations to $\sum_{i=1}^{g}{|\mathcal{A}_i|}$ computations. Following theorem gives a set of sufficient conditions for $g$-group decodability of an STBC.
\begin{theorem}[~\cite{KhR},~\cite{KaR2},~\cite{RaR}] \label{MultigroupTheorem}
An STBC $\mathcal{C}({\bf X},\mathcal{A})$ is $g$-group decodable if there exists a partition of $N_K$ into $g$ non-empty subsets \mbox{$\Gamma_1,\dots,\Gamma_g$} and if there exist finite subsets \mbox{$\mathcal{A}_{i} \subset \mathbb{R}^{|\Gamma_i|}$}, $i=1,\dots,g$ such that the following conditions are satisfied
	\begin{enumerate}
	
	\item The weight matrices $A_l$, $l \in N_K$ of the design ${\bf X}$ are such that
					\begin{equation*}
					A_k^HA_l + A_l^HA_k = {\bf 0} \textrm{ whenever } l \in \Gamma_i , k \in \Gamma_j \textrm{ and } i \neq j 
					\end{equation*}
	\item $\mathcal{C}({\bf X},\mathcal{A})$ is $g$-group encodable with respect to the partition \mbox{$\Gamma_1,\dots,\Gamma_g$} of the set $N_K$ and the signal sets \mbox{$\mathcal{A}_1,\dots,\mathcal{A}_g$}. 	
	\end{enumerate}
\end{theorem}

We now turn to fast-decodable codes. Consider an STBC $\mathcal{C}({\bf X},\mathcal{A})$ being used for communication via the MIMO channel~\eqref{MIMOChannel}. For a complex matrix $A$, define
\begin{equation*}
\tilde{vec}(A) = [vec(A_{Re})^T ~ vec(A_{Im})^T]^T
\end{equation*}
In~\cite{HaH} it was shown that the ML decoding problem can be reduced to one of finding
\begin{equation*}
\textrm{arg min}_{x \in \mathcal{A}} ||y -\mathcal{H}x||^2_F
\end{equation*}
where 
\begin{align}
&y = \tilde{vec}(Y) & x=[x_1,\dots,x_K]^T \nonumber
\end{align}
\begin{equation*}
\mathcal{H} = \begin{bmatrix} \tilde{vec}(A_1H) ~ \tilde{vec}(A_2H) ~ \cdots ~ \tilde{vec}(A_KH) \end{bmatrix}
\end{equation*}
Let the QR decomposition of $\mathcal{H}$ be \mbox{$\mathcal{H} = Q\mathcal{R}$} where $\mathcal{R}$ is a $K \times K$ upper triangular matrix. The ML decoding problem now reduces to
\begin{equation*}
\textrm{arg min}_{x \in \mathcal{A}} ||Q^Ty -\mathcal{R}x||^2_F
\end{equation*}
The $i^{th}$ column of $\mathcal{R}$ is associated with the symbol $x_i$. Consider the case when $\mathcal{R}$ is of the following form
\begin{equation} \label{FastR}
\mathcal{R} = \begin{pmatrix}
								T_1 & {\bf 0} & \ldots & {\bf 0} & B_1 \\
								{\bf 0} & T_2 & \ldots & {\bf 0} & B_2 \\
								\vdots & \vdots & \ldots & \vdots & \vdots \\
								{\bf 0} & {\bf 0} & \ldots & T_g & B_g \\
								{\bf 0} & {\bf 0} & \ldots & {\bf 0} & T_{g+1}
								\end{pmatrix} 
\end{equation}
where $g \geq 1$, $T_i$, $i = 1,\dots,g+1$ are square upper triangular matrices and $B_i$, in general, are rectangular. Denote by $\tilde{\Gamma}_i$ the set of indices of the columns associated with the matrix $T_i$. During sphere-decoding\cite{ViB}, \cite{DCB} when values are proposed for the symbols $\{x_l|l \in \tilde{\Gamma}_{g+1}\}$, the remaining symbols become conditionally $g$-group decodable, provided they have been encoded independently. This reduces the ML decoding complexity greatly. Such an STBC is said to be fast-decodable. Here we call such STBCs as being `conditionally $g$-group decodable'. Formal definition of a fast-decodable STBC is given below.
\begin{definition}
Let $\mathcal{C}({\bf X},\mathcal{A})$ be such that for every channel realization $H$, the upper triangular matrix $\mathcal{R}(H)$ is of the form given in~\eqref{FastR}. If further the tuples $x_{\tilde{\Gamma}_i}$, $i=1,\dots,g$ are encoded independently then $\mathcal{C}({\bf X},\mathcal{A})$ is said to be fast-decodable.
\end{definition}

The following theorem gives the relation between the structure of $\mathcal{R}$ and the linear dispersion matrices of the design ${\bf X}$.
\begin{theorem}[~\cite{SrR1}] \label{PavanTheorem2}
Let $r_{l,k}(H)$ be the element in the $l^{th}$ row and $k^{th}$ column of $\mathcal{R}(H)$. For any $1 \leq l < k \leq K$ and any channel realization $H$, $r_{l,k}(H) = 0$ if \mbox{$A_k^HA_l + A_l^HA_k = {\bf 0}$}.
\end{theorem}

This means that an STBC $\mathcal{C}({\bf X},\mathcal{A})$ will be conditionally $g$-group decodable if the STBC obtained by suppressing a subset of the variables is $g$-group decodable. More formally we have the following theorem which gives a sufficient condition for an STBC to be fast-decodable.
\begin{theorem} \label{FDSufficientCondition}
Consider a $\lambda$-group encodable STBC $\mathcal{C}({\bf X},\mathcal{A})$ with the encoding groups being $\Gamma_1,\dots,\Gamma_{\lambda}$. $\mathcal{C}({\bf X},\mathcal{A})$ is fast-decodable if $\tilde{\Gamma}_1,\dots,\tilde{\Gamma}_g \subseteq N_K$ are such that 
we have
\begin{equation} \label{FDTheorem1}
A_k^HA_l + A_l^HA_k = {\bf 0}  
\end{equation}
whenever
\begin{equation} \label{FDTheorem2}
l \in \Gamma_p , k \in \Gamma_q \textrm{ and } p \neq q.
\end{equation}
\end{theorem}
\begin{proof}
From Theorem~\ref{PavanTheorem2} and~\eqref{FDTheorem1} it is clear that $r_{l,k}(H) = 0$ for any channel realization $H$ whenever $l \in \tilde{\Gamma}_i , k \in \tilde{\Gamma}_j \textrm{ and } i \neq j$. Thus with appropriate reordering of the columns the matrix $\mathcal{R}(H)$ can be brought to the form in~\eqref{FastR}. It only remains to show that the tuples $x_{\tilde{\Gamma}_i}$, $i=1,\dots,g$ are encoded independently. But~\eqref{FDTheorem2} assures exactly this. Thus $\mathcal{C}({\bf X},\mathcal{A})$ is fast-decodable.
\end{proof}

Fast-Group-Decodable(FGD) STBCs were recently introduced in~\cite{RGYS}. 
\begin{definition}[~\cite{RGYS}] \label{FGDDefinition}
An STBC $\mathcal{C}({\bf X},\mathcal{A})$ is said to be fast-group-decodable if it satisfies the following conditions
	\begin{enumerate}
	
	\item For some $g>1$, $\mathcal{C}({\bf X},\mathcal{A})$ is $g$-group decodable with the groups and the signal sets as $\Gamma_1,\dots,\Gamma_g$ and $\mathcal{A}_1,\dots,\mathcal{A}_g$.
	
	\item There exists an $i \in \{1,\dots,g\}$ such that the STBC $\mathcal{C}({\bf X_{\Gamma_i}},\mathcal{A}_i)$ is fast-decodable.
	
	\end{enumerate}
\end{definition}

The complexity is reduced because of two reasons. Firstly multigroup decodability decreases the number of computations. Further there is at least one component code which is fast-decodable. Such a low complexity code was given in~\cite{RGYS} for $4$ transmit antennas and rate $17/8$ cspcu. 

So far we have only discussed about the decoding complexity of an STBC $\mathcal{C}({\bf X},\mathcal{A})$. We end this section by introducing the notion of the decoding complexity of a design ${\bf X}$. From Theorem~\ref{MultigroupTheorem} it is clear that the decoding complexity of an STBC $\mathcal{C}({\bf X},\mathcal{A})$ is affected by the choice of both the design ${\bf X}$ and the signal set $\mathcal{A}$. By the decoding complexity of a design we refer to the amount of complexity that the choice of the linear dispersion matrices contributes to the decoding complexity of the STBC. The following definition makes this formal. 
\begin{definition}
Consider a design ${\bf X} = \sum_{i=1}^{K}{x_iA_i}$.
\begin{enumerate} 
	\item {\bf X} is said to be $g$-group decodable if there exists a partition of $N_K$ into $g$ non-empty subsets \mbox{$\Gamma_1,\dots,\Gamma_g$} such that
					\begin{equation}
					A_k^HA_l + A_l^HA_k = {\bf 0} \textrm{ whenever } l \in \Gamma_i , k \in \Gamma_j \textrm{ and } i \neq j 					\label{A_antiHcomm}
					\end{equation}
	\item ${\bf X}$ is said to be fast-decodable if there exists a non-empty subset $\Gamma \subseteq N_K$ such that the design ${\bf X_{\Gamma}}$ is $g$-group decodable for some $g>1$.
	\item ${\bf X}$ is said to be fast-group-decodable if ${\bf X}$ is $g$-group decodable with respect to the partition \mbox{$\Gamma_1,\dots,\Gamma_g$} and there exists at least one $\Gamma_i$ such that ${\bf X_{\Gamma_i}}$ is fast-decodable.
\end{enumerate}
\end{definition}


\section{On Full Diversity} \label{FullDiversitySection}

In this section we give a sufficient condition for a design ${\bf X}$ to give rise to full-diversity STBC $\mathcal{C}({\bf X},\mathcal{A})$ via single real symbol encoding. This result is then used to show that every low complexity design presented in this work can be combined with an appropriate signal set to get a full diversity STBC without compromising on decoding comfort. 


Let ${\bf X_n} = \sum_{i=1}^{n}x_iA_i$ be an $N \times N$ linear design in $n$ real symbols $\{x_1,\dots,x_n\}$ and let \mbox{$A_i \in \mathbb{C}^{N \times N}$}, $i=1,\dots,n$ be full-rank. Given a set of $n$ positive integers $Q_i$, $i=1,\dots,n$ we are interested in finding a real constellation $\mathcal{A}_i \subset \mathbb{R}$ for the real symbol $x_i$ with $|\mathcal{A}_i|=Q_i$ for each $i=1,\dots,n$. The constellations must be such that the specific STBC obtained, \mbox{$\mathcal{C}({\bf X_n},\mathcal{A}_1 \times \dots \times \mathcal{A}_n)$}, must be of full-diversity. Towards establishing the main result of this section we introduce some notation.

For each $i \in \{1,\dots,n\}$ let \mbox{$\mathcal{A}_i = \{a_i[0],a_i[1],\dots,a_i[Q_i-1]\}$}, where $a_i[j] \in \mathbb{R}$. For an $n$ tuple \mbox{$u = (u_1,\dots,u_n) \in \mathbb{Z}_{Q_1} \times \mathbb{Z}_{Q_2} \dots \times \mathbb{Z}_{Q_n}$} define 
\begin{align}
C_n[u] &= {\bf X_n}(a_1[u_1],a_2[u_2],\dots,a_n[u_n]) \nonumber \\
			 &= \sum_{i=1}^{n}{a_i[u_i]A_i}	
\end{align}
Hence the single real symbol encodable STBC obtained by using the given constellations satisfies
\begin{equation*}
\mathcal{C}({\bf X_n},\mathcal{A}_1 \times \dots \times \mathcal{A}_n) = \{C_n[u]|u \in \mathbb{Z}_{Q_1} \times \dots \times \mathbb{Z}_{Q_n}\}
\end{equation*}
We see that the codewords are indexed by the elements of \mbox{$\mathbb{Z}_{Q_1} \times \dots \times \mathbb{Z}_{Q_n}$}. 

\begin{theorem} \label{InductionStep}
Let ${\bf X_n} = \sum_{i=1}^{n}{x_iA_i}$ be an $N \times N$ linear design in $n$ real variables with full-rank weight matrices $A_i$. Let $\mathcal{A}_i \subset \mathbb{R}$, $i=1,\dots,n$ be such that $|\mathcal{A}_i| = Q_i$ and \mbox{$\mathcal{C}({\bf X_n},\mathcal{A}_1 \times \dots \times \mathcal{A}_n)$} is of full diversity. Let \mbox{$A_{n+1} \in \mathbb{C}^{N \times N}$} be any full rank matrix and $Q_{n+1}$ be any positive integer. Then there exists a one dimensional real constellation \mbox{$\mathcal{A}_{n+1} \subset \mathbb{R}$} such that
	\begin{enumerate}
	\item $|\mathcal{A}_{n+1}| = Q_{n+1}$
	\item The STBC \mbox{$\mathcal{C}({\bf X_{n+1}},\mathcal{A}_1 \times \dots \times \mathcal{A}_{n+1})$} offers full diversity.
	\end{enumerate}
\end{theorem}
\begin{proof}
Proof is given in Appendix A.
\end{proof}

We now present the main result of this section in the following theorem.
\begin{theorem} \label{FullDiversityTheorem}
For any given $N \times N$ square linear design ${\bf X_n} = \sum_{i=1}^{n}{x_iA_i}$ with full-rank weight matrices $A_i$ and positive integers $Q_1,\dots,Q_n$, there exist constellations $\mathcal{A}_i \subset \mathbb{R}$, $i=1,\dots,n$ such that
\begin{enumerate}
\item $|\mathcal{A}_i| = Q_i$ for $i=1,\dots,n$
\item The single real symbol encodable STBC \mbox{$\mathcal{C}({\bf X_n},\mathcal{A}_1 \times \dots \times \mathcal{A}_n)$} offers full diversity.
\end{enumerate}
\end{theorem}
\begin{proof}
Proof is by induction. The theorem is shown to be true for $n=1$ here. Theorem~\ref{InductionStep} is the induction step.

Consider the design for one real symbol ${\bf X_1} = x_1A_1$. Choose any \mbox{$\mathcal{A}_1 \subset \mathbb{R}$} with \mbox{$|\mathcal{A}_1| = Q_1$}. The codewords are indexed by elements in $\mathbb{Z}_{Q_1}$. For any \mbox{$u,v \in \mathbb{Z}_{Q_1}$} and $u \neq v$ we have 
\begin{align}
det(C_1[u] - C_1[v]) &= det((a_1[u]-a_1[v])A_1) \nonumber \\
										 &= (a_1[u]-a_1[v])^Ndet(A_1) \nonumber \\
										 &\neq 0 
\end{align}
Since the difference matrix of any two codewords is of full-rank, the STBC $\mathcal{C}({\bf X_1},\mathcal{A}_1)$	offers full diversity.									
\end{proof}

The STBC obtained from Theorem~\ref{FullDiversityTheorem} is single real symbol encodable. Thus if ${\bf X}$ were a $g$-group decodable, fast-decodable or fast-group-decodable design then the resulting STBC $\mathcal{C}({\bf X},\mathcal{A})$ is a $g$-group decodable, fast-decodable or fast-group-decodable STBC respectively. Most importantly the condition that the linear dispersion matrices are full-rank ensures that the STBC $\mathcal{C}({\bf X},\mathcal{A})$ offers full-diversity. 

All the designs discussed in this paper have unitary, and hence, full-rank weight matrices. In the remaining sections of the paper the focus is on designing low complexity designs and not on the design of signal sets $\mathcal{A}$. Since it has been shown in Theorem~\ref{FullDiversityTheorem} that there exist signal sets leading to full-diversity STBCs without increasing the decoding complexity than that imposed by the design itself, the only problem left is to design signal sets that maximize the coding gain without increasing the decoding complexity. This problem is not addressed in this paper.

The following theorem will be useful when we are constructing STBCs with low decoding complexity.
\begin{theorem} \label{RegularPAMTheorem}
Let ${\bf X} = \sum_{i=1}^{K}{x_iA_i}$ be an $N \times N$ design in $K$ real symbols with full-rank weight matrices and $l \in \{1,\dots,K\}$ be such that
	\begin{equation}
	A_i^HA_j + A_j^HA_i = 2\delta_{i,j}I_N \textrm{ for all } 1 \leq i,j \leq l \label{HurwitzRadon}
	\end{equation}
Given positive integers $Q_1,\dots,Q_K$ and any set of real constellations $\mathcal{A}_1,\dots,\mathcal{A}_l$ with cardinalities $Q_1,\dots,Q_l$ respectively, there exist real constellations $\mathcal{A}_i$, $l < i \leq K$ such that
	\begin{enumerate}
	\item The STBC $\mathcal{C}({\bf X},\mathcal{A}_1\times \dots \times \mathcal{A}_K)$ offers full diversity.
	\item $|\mathcal{A}_i| = Q_i$ for $l < i \leq K$.
	\end{enumerate}
\end{theorem}
\begin{proof}
Consider the design ${\bf X_l} = \sum_{i=1}^{l}{x_iA_i}$ and the STBC $\mathcal{C}({\bf X_l},\mathcal{A}_1\times \dots \times \mathcal{A}_l)$ generated using the signal sets $\mathcal{A}_1,\dots,\mathcal{A}_l$ for the independent variables $x_1,\dots,x_l$. Because $\{A_1,\dots,A_l\}$ satisfy the complex Hurwitz-Radon matrix equations \eqref{HurwitzRadon} we have \cite{Lia},
\begin{equation*}
{\bf X_l}^H{\bf X_l} = \sum_{i=1}^{l}{x_i^2A_i^HA_i} = (\sum_{i=1}^{l}{x_i^2})I_N
\end{equation*}
Thus for any real signal sets $\mathcal{A}_1,\dots,\mathcal{A}_l$, and \mbox{$u,v \in \mathbb{Z}_{Q_1} \times \dots \times \mathbb{Z}_{Q_l}$} with $u \neq v$ we have
\begin{equation*}
(C_l[u] - C_l[v])^H(C_l[u] - C_l[v]) = \sum_{i=1}^{l}{(a_i[u_i]-a_i[v_i])^2}I_N 
\end{equation*}
which is full-rank. Since the rank of any square matrix $A$ is equal to the rank of $A^HA$, $det(C_l[u] - C_l[v]) \neq 0$. Thus the STBC $\mathcal{C}({\bf X_l},\mathcal{A}_1\times \dots \times \mathcal{A}_l)$ is of full diversity. By using Theorem~\ref{InductionStep} repeatedly $K-l$ times with integers $Q_i$ and the matrices $A_i$, $l<i \leq K$ we get the desired signal set and the STBC.
\end{proof}


\section{Low Decoding Complexity STBCs via codes over $\mathbb{F}_4$} \label{CliffordAlgebrasSection}

In this section, we make use of Universal Clifford Algebras and their matrix representations over the complex field, to construct low decoding complexity codes via codes over $\mathbb{F}_4.$  A subset of elements in the Universal Clifford Algebra are seen to have multiplicative properties similar to~\eqref{Hurwitz-Radon-orthogonality}. We attempt to obtain low complexity designs by choosing linear dispersion matrices from the matrix representation of these elements. We proceed in this direction by using a theorem that establishes an isomorphism between a Universal Clifford Algebra and a full matrix algebra of appropriate dimension over $\mathbb{C}$. It is then observed that the tensor product of Pauli matrices are a double cover of the set elements in question from the Universal Clifford Algebra. The connection with vectors over $\mathbb{F}_4$ is made. The problem of finding low complexity designs is then converted to one of finding a set of vectors in $\mathbb{F}_2 \oplus \mathbb{F}^m_4$. Examples of low complexity STBCs available in the literature that are obtainable from codes over $\mathbb{F}_4$ are given.

Let $n$ be any positive integer. Denote by $N_n$ the set $\lbrace 1,\dots,n \rbrace$. Let \mbox{$e_1,\dots,e_n$} be elements of an associative algebra over $\mathbb{C}$ and $\alpha \subseteq N_n$. For any nonempty subset \mbox{$\alpha = \lbrace i_1,\dots,i_{\vert\alpha\vert}\rbrace$} with \mbox{$i_1 < i_2 < \cdots < i_{\vert\alpha\vert }$} we define \mbox{$e_{\alpha} = e_{i_1}e_{i_2} \cdots e_{i_{\vert\alpha\vert}}$} and \mbox{$e_{\phi} = 1$}.

\begin{definition}[\cite{GiM}]
A Universal Clifford Algebra $\mathcal{U}_n$ is an associative algebra over $\mathbb{C}$ with a multiplicative identity 1 and generated by $n$ objects $e_1,\dots,e_n$ which satisfy the following equations
	\begin{equation}
	e_ie_j + e_je_i = 0 \textrm{ whenever } i \neq j \label{e_anticomm}
	\end{equation}
	\begin{equation}
	{e_i}^2 = -1 \textrm{ for } i = 1,\dots,n \label{e_square}
	\end{equation}
	\begin{equation}
	\lbrace e_{\alpha} \vert \alpha \subseteq N_n \rbrace \textrm{ is a basis for } \mathcal{U}_n \label{e_basis}
	\end{equation}
\end{definition}

From \eqref{e_anticomm} and \eqref{e_square} it is clear that for any $\alpha \subseteq N_n$, ${e_{\alpha}}^2 = \pm1$. Also for any $\alpha,\beta \subseteq N_n$ either \mbox{$e_{\alpha}e_{\beta} +  e_{\beta}e_{\alpha} = 0$} or \mbox{$e_{\alpha}e_{\beta} - e_{\beta}e_{\alpha} = 0$}. This property resembles \eqref{Hurwitz-Radon-orthogonality} except for the conjugate-transpose. Hence by representing the basis elements using either Hermitian or skew-Hermitian matrices we can obtain linear dispersion matrices that are Hurwitz-Radon orthogonal. Together with the fact that ${e_{\alpha}}^2 = \pm1$ it is clear that we need unitary representation of the basis elements. The following theorem gives a representation of a class of Clifford Algebras.
\begin{theorem}[~\cite{GiM}] \label{Representation}
For any positive integer $m$, the Clifford Algebra $\mathcal{U}_{2m}$ is isomorphic to the full matrix algebra $\mathbb{C}^{2^m \times 2^m}$. The extension of the map \mbox{$e_k \rightarrow E_k \textrm{ for } k = 1,\dots,2m$} gives an isomorphism of the algebras, where for $s = 1,\dots,m$, $E_s$ and $E_{s+m}$ are the $m$-fold tensor products given by
	\begin{equation*}
	E_s = i(Z \otimes \cdots \otimes Z \otimes iXZ \otimes I_2 \otimes \cdots \otimes I_2)
	\end{equation*}
	\begin{equation*}
	E_{s+m} = i(Z \otimes \cdots \otimes Z \otimes X \otimes I_2 \otimes \cdots \otimes I_2)
	\end{equation*}
there being $s-1$ factors of $Z$ in each product.
\end{theorem}

From Theorem~\ref{Representation} we have that each $E_k$, $k = 1,\dots,2m$ is unitary, skew-Hermitian and thus squares to $-I$. With $E_{\alpha}$ defined similar to $e_{\alpha}$ for \mbox{$\alpha \subseteq N_{2m}$} we see that all the basis elements are represented in terms of unitary matrices. The isomorphism ensures that these matrices are linearly independent over $\mathbb{C}$. Since we are concerned with the transmission of real symbols we note that $\mathcal{B} = \lbrace i^{\lambda}E_{\alpha} \vert \lambda \in \mathbb{Z}_2 \textrm{ and } \alpha \subseteq N_{2m} \rbrace$ is a $\mathbb{R}$-linear basis for $\mathbb{C}^{2^m \times 2^m}$. With $-\mathcal{B}$ defined as \mbox{$\{-b|b \in \mathcal{B}\}$}, we have the following proposition.
\begin{proposition} \label{BandG}
$\mathcal{B} \cup -\mathcal{B} = G_m$.
\end{proposition}
\begin{proof}
We note that both $X$ and $Z$ square to $I_2$. Further, they anticommute. So it is clear that for any $\alpha \subseteq N_{2m}$, $E_{\alpha} \in G_m$. Since $G_m$ is closed under multiplication by $iI_{2^m}$ and $-I_{2^m}$ we have \mbox{$\mathcal{B} \cup -\mathcal{B}$ $\subseteq G_m$}. To complete the proof we note there are $2^{2m}$ distinct subsets of $N_{2m}$, thus $|\mathcal{B}| = 2^{2m+1}$. Since $\mathcal{B}$ is linearly independent over $\mathbb{R}$, for any \mbox{$b \in \mathcal{B}$} we have $-b \notin \mathcal{B}$. Equivalently \mbox{$\mathcal{B} \cap -\mathcal{B} = \phi$}. Thus, \mbox{$|\mathcal{B} \cup -\mathcal{B}| = 2^{2m+2} = |G_m|$}. Thus \mbox{$\mathcal{B} \cup -\mathcal{B} = G_m$}.
\end{proof}

Proposition~\ref{BandG} says that the weight matrices to be chosen from the matrix representation of the basis elements of Clifford Algebra can be equivalently obtained through the Pauli group. However, the Pauli group of matrices are not linearly independent over $\mathbb{R}$. Thus we concern ourselves with a proper subset $\Lambda_m$ of $G_m$ which is maximally linearly independent and thus satisfies \mbox{$\Lambda_m \cup -\Lambda_m = G_m$}. One such set is the following
\begin{equation}
\Lambda_m = \lbrace ~ i^{\lambda}B_1 \otimes \cdots \otimes B_{m} \vert \lambda \in \mathbb{Z}_2 \textrm{ and } B_k \in \{I_2,iX,iZ,ZX \} ~ \rbrace
\end{equation}
\begin{proposition}
The set $\Lambda_m$ is a basis for $\mathbb{C}^{2^m \times 2^m}$ as a vector space over $\mathbb{R}$.
\end{proposition}
\begin{proof}
$\{I_2,iX,iZ,ZX \}$ is a basis for $\mathbb{C}^{2 \times 2}$ as a vector space over $\mathbb{C}$. Thus their $m$ fold tensor products form a $\mathbb{C}$-linear basis for $\mathbb{C}^{2^m \times 2^m}$. From this the required result follows.
\end{proof}

We now proceed by relating the set $\Lambda_m$ to $\mathbb{F}_2 \oplus \mathbb{F}^m_4$. Consider the finite field $\mathbb{F}_4$ with $4$ elements \mbox{$\lbrace 0,1,\omega,\omega^2 \rbrace$} satisfying the relations \mbox{$1 + \omega = \omega^2.$}  Define a map 
\begin{equation*}
\psi : \{I_2,iX,iZ,ZX\} \to \mathbb{F}_4
\end{equation*}
that sends 
\begin{align}
& I_2 \to 0 \textrm{, } iX \to 1 \textrm{, } iZ \to \omega \textrm{, } ZX \to \omega^2
\end{align}
Note that every element \mbox{$t \in \Lambda_m$} can be uniquely written as
\begin{equation}
t = i^\lambda\psi^{-1}(\xi_1) \otimes \cdots \otimes \psi^{-1}(\xi_m) \textrm{ for } \lambda \in \mathbb{Z}_2 \textrm{ and } \xi_k \in \mathbb{F}_4
\end{equation}
\begin{proposition}
The map $\varphi : \Lambda_m \to \mathbb{F}_2 \oplus \mathbb{F}_4^m$ that sends
	\begin{equation*}
	i^{\lambda}B_1 \otimes \cdots \otimes B_{m} \to [\lambda,\psi(B_1),\dots,\psi(B_m)]
	\end{equation*}
is a one to one correspondence between $\Lambda_m$ and $\mathbb{F}_2 \oplus \mathbb{F}_4^m$.
\end{proposition}
\begin{proof}
Since $\psi$ is one to one, it is clear that $\varphi$ is one to one. Further $|\Lambda_m| = |\mathbb{F}_2 \oplus \mathbb{F}_4^m| = 2^{2m+1}$. Thus $\psi$ is surjective as well. This completes the proof.
\end{proof}

The Hamming weight of a vector in $\mathbb{F}_2 \oplus \mathbb{F}_4^m$ is defined next.
\begin{definition}
The (Hamming) weight ${\sf wt}([\lambda,\xi_1,\dots,\xi_m])$ of a vector \mbox{$[\lambda,\xi_1,\dots,\xi_m] \in \mathbb{F}_2 \oplus \mathbb{F}_4^m$} is defined as
	\begin{equation*}
	{\sf wt}([\lambda,\xi_1,\dots,\xi_m]) = {\sf \bf 1}\{ \lambda \neq 0 \} + \sum_{i=1}^{m}{{\sf \bf 1}\{ \xi_i \neq 0 \}}
	\end{equation*}
\end{definition}

Any matrix $t \in \Lambda_m$ is either Hermitian or skew-Hermitian. This information about $t$ is present in ${\sf wt}(\varphi(t))$. The following proposition explains this claim.
\begin{proposition} \label{SHcondn}
A matrix $t \in \Lambda_m$ is Hermitian if ${\sf wt}(\varphi(t))$ is even. Else it is skew-Hermitian.
\end{proposition}
\begin{proof}
Let \mbox{$t = i^{\lambda}B_1 \otimes \cdots \otimes B_{m}$}, $B_i \in \{I_2,iX,iZ,ZX\}$ and $\lambda \in \mathbb{Z}_2$. Except $I_2$,which is mapped to $0$ under $\psi$, the other $3$ matrices are skew-Hermitian. Hence
\begin{align}
t^{H} &= (-1)^{{\sf \bf 1}\{\lambda \neq 0\}}i^{\lambda}(-1)^{{\sf \bf 1}\{B_1 \neq I\}}B_1 \otimes \cdots \otimes (-1)^{{\sf \bf 1}\{B_m \neq I\}}B_m \nonumber \\
			&= (-1)^{{\sf \bf 1}\{\lambda \neq 0\}}i^{\lambda} \otimes_{k=1}^{m}{(-1)^{{\sf \bf 1}\{\psi(B_k) \neq 0\}}B_k} \nonumber \\
			&= (-1)^{{\sf wt}(\varphi(t))}t \nonumber
\end{align}
Thus $t$ is Hermitian if ${{\sf wt}(\varphi(t))}$ is even, else $t$ is skew-Hermitian.
\end{proof}

For linear dispersion matrices coming from the set $\Lambda_m$ the $g$-group ML decodability condition \eqref{A_antiHcomm} can be reformulated in terms of the weight of corresponding vectors in $\mathbb{F}_2 \oplus \mathbb{F}_4^m$ as follows: 
\begin{proposition} \label{t_antiHcomm}
For any $t_1,t_2 \in \Lambda_m$, we have
	\begin{equation*}
	t_1^Ht_2 + t_2^Ht_1 = {\bf 0} \textrm{~ iff ~} {\sf wt}(\varphi(t_1) + \varphi(t_2)) \textrm{ is odd} 
	\end{equation*}
where the vector sum is component wise addition.
\end{proposition}
\begin{proof}
Recall that any \mbox{$t \in \Lambda_m$} is either Hermitian or skew-Hermitian. Thus $t_1^Ht_2$ is skew-Hermitian iff $t_1t_2$ is skew-Hermitian. Let
	\begin{equation*}
	t_k = i^{\lambda_k}\psi^{-1}(\xi_{k,1}) \otimes \cdots \otimes \psi^{-1}(\xi_{k,m}) \textrm{ for } k = 1,2
	\end{equation*}
Note that \mbox{$\psi^{-1}(\zeta)\psi^{-1}(\eta) = \pm \psi^{-1}(\zeta + \eta)$} for any \mbox{$\zeta,\eta \in \mathbb{F}_4$}. Hence $t_1t_2$
	\begin{align}
	&= \pm i^{\lambda_1+\lambda_2}\psi^{-1}(\xi_{1,1}+\xi_{2,1}) \otimes \cdots \otimes \psi^{-1}(\xi_{1,m}+\xi_{2,m}) \nonumber \\
				 &= \pm i^{(\lambda_1+\lambda_2) \textrm{mod}2}\psi^{-1}(\xi_{1,1}+\xi_{2,1}) \otimes \cdots \otimes \psi^{-1}(\xi_{1,m}+\xi_{2,m}) \nonumber \\
				 &= \pm \varphi^{-1}(\varphi(t_1) + \varphi(t_2)) \nonumber
	\end{align}
We have $t_1t_2$ is skew-Hermitian iff \mbox{$\varphi^{-1}(\varphi(t_1) + \varphi(t2))$} is skew-Hermitian. Applying Proposition~\ref{SHcondn} we have that $t_1t_2$ is skew-Hermitian iff ${\sf wt}(\varphi(t_1) + \varphi(t_2))$ is odd. This completes the proof.
\end{proof}

When we restrict the possible linear dispersion matrices to the set $\Lambda_m$, Proposition~\ref{t_antiHcomm} helps us reformulate the original problem of finding weight matrices for low decoding complexity STBCs in terms of finding vectors in $\mathbb{F}_2 \oplus \mathbb{F}_4^m$. This leads to the following theorem.
\begin{theorem} \label{Main}
If there exist $K$ distinct vectors \mbox{$y_1,\dots,y_K \in \mathbb{F}_2 \oplus \mathbb{F}_4^m$} and a partition  \mbox{$\Gamma_1,\dots,\Gamma_g$} of $\{1,\dots,K\}$ into nonempty subsets such that
	\begin{equation}
	{\sf wt}(y_k + y_l) \textrm{ is odd } \textrm{ whenever } k \in \Gamma_i \textrm{, } l \in \Gamma_j \textrm{ and } i \neq j
	\end{equation}
then there exists a design $\mathbf{X}(x_1,\dots,x_K)$ of dimension \mbox{$2^m \times 2^m$} in $K$ real variables and which is $g$-group ML decodable with the $i^{th}$ group being \mbox{$\{ x_k \vert k \in \Gamma_i \}$}.
\end{theorem}
\begin{proof}
Given the $K$ vectors as in the hypothesis, define \mbox{$A_k = \varphi^{-1}(y_k) \in \Lambda_m$}. The bijective nature of $\varphi$ ensures that the $K$ matrices $A_k$ are distinct. Since $\Lambda_m$ is linearly independent over $\mathbb{R}$, $A_k$,$k=1,\dots,K$ are linearly independent. Define a linear space-time design $\mathbf{X}(x_1,\dots,x_K)$ as
	\begin{equation*}
	\mathbf{X} = \sum_{i=1}^{K}{x_iA_i}
	\end{equation*}
Applying Proposition~\ref{t_antiHcomm} we get the $i^{th}$ group as $\Gamma_i$. This completes the proof.
\end{proof}

Theorem~\ref{Main} converts the original problem of finding $g$-group ML decodable designs from $\Lambda_m$ to that of finding certain codes over $\mathbb{F}_4$. Once such a code  is chosen in \mbox{$\mathbb{F}_2 \oplus \mathbb{F}_4^m$} the linear dispersion matrices can be obtained by the one-to-one correspondence $\varphi$. 
\begin{definition}
A design in $K$ real symbols for $2^m$ antennas from \mbox{$\mathbb{F}_2 \oplus \mathbb{F}_4^m$} is defined as a subset \mbox{$\mathcal{S} \subseteq \mathbb{F}_2 \oplus \mathbb{F}_4^m$} such that \mbox{$|\mathcal{S}|=K$}.
\end{definition}

The `design' $\mathcal{S}$ in the above definition refers to the linear design that can be obtained by mapping the vectors in \mbox{$\mathcal{S}=\{y_1,\dots,y_K\}$} to linear dispersion matrices in $\Lambda_m$. The corresponding design in $K$ real variables \mbox{$x_1,\dots,x_K$} is given as
\begin{equation}
{\bf X} = \sum_{i=1}^{K}{x_i\varphi^{-1}(y_i)} \label{StoX}
\end{equation}
With the linear design ${\bf X}$ associated with the set $\mathcal{S}$ defined as in~\eqref{StoX} we can use Theorem~\ref{Main} to define $g$-group decodable, FD and FGD designs obtainable from codes over $\mathbb{F}_4$. These are given below.
\begin{definition} \label{LowComplexityDefinition}
Let $\mathcal{S}$ be a design obtainable from \mbox{$\mathbb{F}_2 \oplus \mathbb{F}_4^m$}.
\begin{enumerate}
\item \mbox{$\mathcal{S}=\cup_{i=1}^{g}{\mathcal{S}_i}$} or equivalently the set \mbox{$\{\mathcal{S}_k | k = 1,\dots,g\}$} is called a $g$-group decodable design if for any $y \in \mathcal{S}_k$, $z \in \mathcal{S}_l$ and $k \neq l$ we have ${\sf wt}(y + z)$ is odd.
\item $\mathcal{S}$ is said to be fast-decodable if there exist subsets \mbox{$\mathcal{S}_i \subseteq \mathcal{S}$}, $i=1,\dots,g$, $g > 1$ such that \mbox{$\{\mathcal{S}_i|i=1,\dots,g\}$} is a $g$-group decodable design.
\item A $g$-group decodable design \mbox{$\{ \mathcal{S}_i | i=1,\dots,g \}$} is said to be fast-group-decodable if there exists at least one \mbox{$l \in \{1,\dots,g\}$} such that the design $\mathcal{S}_l$ is fast-decodable.
\end{enumerate}
\end{definition} 

We now give examples of low complexity designs obtainable from codes over $\mathbb{F}_4$ i.e., linear designs whose weight matrices come from $\Lambda_m$. Assume that the number of groups is $g$ and each group has $\tau$ vectors in it. The total number of vectors or the total number of real symbols in the design is thus \mbox{$K=g\tau$}.

\subsection{Alamouti Code}
The Alamouti Code~\cite{Ala} is a $2 \times 2$ square orthogonal design of rate $1$. Its parameters are: $m=1$, $K=4$, $g=4$ and $\tau=1$. Its linear dispersion matrices are : \mbox{$\{ I_2,iX,iZ,ZX \}$}. All the weight matrices belong to $\Lambda_1$. The four sets of vectors from \mbox{$\mathbb{F}_2 \oplus \mathbb{F}_4$} corresponding to the four groups are as follows:
\begin{equation*}
\mathcal{S}_1 = \{{[0,0]}\} \textrm{, }\mathcal{S}_2 = \{{[0,1]}\} \textrm{, }\mathcal{S}_3 = \{{[0,\omega]}\} \textrm{, }\mathcal{S}_4 = \{{[0,\omega^2]}\}
\end{equation*}
It can be seen that the weight of the sum of any two different vectors is odd, thus the above design is single real symbol decodable.

\subsection{Other $2 \times 2$ codes of rate $1$}
Here we describe designs with parameters $m = 1$, $K = 4$, $g = 2$, $\tau = 2$, $R = 1$. There are only three non-equivalent designs that can be obtained from $\Lambda_1$. They are parametrized by $l \in \{0,1,2\}$ and are given by
\begin{equation*}
\mathcal{S}_1 = \{{[0,0],[1,\omega^l]}\} 
\end{equation*}
\begin{equation*}
\mathcal{S}_2 = \{{[0,\omega^l],[1,0]}\}
\end{equation*}

\subsubsection{$l=0$}
The design is given by \mbox{$\mathcal{S}_1 = \{{[0,0],[1,1]}\}$}, \mbox{$\mathcal{S}_2 = \{{[0,1],[1,0]}\}$}. The two groups of weight matrices are \mbox{$\mathcal{M}_1 = \{{I,-X}\}$}, and \mbox{$\mathcal{M}_2 = \{{iX,iI}\}$}. With $\Gamma_1 = \{1,2\}$ and $\Gamma_2 = \{3,4\}$ the resulting design is
\begin{equation*}
{\bf X} = \begin{pmatrix}
x_1 + ix_4 & -x_2 + ix_3 \\
-x_2 + ix_3 & x_1 + ix_4
\end{pmatrix}
\end{equation*}
This is the $2 \times 2$ ABBA code \cite{TBH}.

\subsubsection{$l=1$}
The design is given by \mbox{$\mathcal{S}_1 = \{{[0,0],[1,\omega]}\}$}, \mbox{$\mathcal{S}_2 = \{{[0,\omega],[1,0]}\}$}. The matrices are \mbox{$\mathcal{M}_1 = \{{I,-Z}\}$ and $\mathcal{M}_2 = \{{iZ,iI}\}$}. With $\Gamma_1 = \{1,2\}$ and $\Gamma_2 = \{3,4\}$  the design is given by
\begin{equation*}
{\bf X}  = \begin{pmatrix}
x_1 - x_2 + i(x_4 + x_3) & 0 \\ 0 & x_1 + x_2 + i(x_4 - x_3)
\end{pmatrix}
\end{equation*}
With appropriate transformation of the symbols within each group we get the $2 \times 2$ CIOD \cite{KhR}.
\begin{equation*}
{\bf X}  = \begin{pmatrix}
\tilde{x}_1 + i\tilde{x}_3 & 0 \\ 0 & \tilde{x}_2 + i\tilde{x}_4
\end{pmatrix}
\end{equation*}

\subsubsection{$l=2$}
\mbox{$\mathcal{S}_1 = \{{[0,0],[1,\omega^2]}\}$} and \mbox{$\mathcal{S}_2 = \{[1,0],[0,\omega^2]\}$}. The linear dispersion matrices are \mbox{$\mathcal{M}_1 = \{{I,iZX}\}$ and $\mathcal{M}_2 = \{{iI,ZX}\}$}. With $\Gamma_1 = \{1,2\}$ and $\Gamma_2 = \{3,4\}$ the resulting design is
\begin{equation*}
{\bf X} = \begin{pmatrix}
x_1 + ix_3 & x_4 + ix_2 \\
-x_4 - ix_2 & x_1 + ix_3
\end{pmatrix}
\end{equation*}
This is the $\begin{pmatrix} a & b \\ -b & a \end{pmatrix}$ design.

\subsection{$4 \times 4$ Quasi-orthogonal design}
Consider the rate $1$ quasi-orthogonal design given in~\cite{Jaf} for $4$ transmit antennas. The design contains $8$ real symbols \mbox{$x_1,\dots,x_8$} and is $4$-group decodable. The parameters are $m=2$, $K=8$, $g=4$, $\tau=2$ and $R=1$. The design \mbox{${\bf X} = \sum_{i=1}^{8}{x_iA_i}$} is given below.
\begin{equation*}
{\bf X} = \begin{pmatrix}
					x_1 +ix_2 &	x_3 +ix_4 &	x_5 +ix_6 &	x_7 +ix_8 \\
					-x_3 +ix_4  &	x_1 -ix_2 &	-x_7 +ix_8 &	x_5 -ix_6 \\
					-x_5 +ix_6 &	-x_7 +ix_8  &	x_1 -ix_2 &	x_3 -ix_4  \\
					x_7 +ix_8 &	-x_5 -ix_6 &	-x_3 -ix_4  &	x_1 +ix_2 
					\end{pmatrix}	
\end{equation*}
The linear dispersion matrices, upto a sign change, $A_i$, $i=1,\dots,8$ are 
\begin{equation*}
\begin{array}{ll}
A_1 = I_2 \otimes I_2 & A_2 = iZ \otimes Z  \\
A_3 = I_2 \otimes ZX  & A_4 = iZ \otimes X  \\
A_5 = ZX \otimes I_2  & A_6 = iX \otimes Z  \\
A_7 = ZX \otimes ZX   & A_8 = iX \otimes X  
\end{array}
\end{equation*}
It can be seen that all the $8$ matrices come from $\Lambda_2$. The corresponding vectors in \mbox{$\mathbb{F}_2 \oplus \mathbb{F}_4^m$}, \mbox{$y_i=\varphi(A_i)$} are
\begin{equation*}
\begin{array}{ll}
y_1 = [0,0,0]									& y_2 = [1,\omega,\omega] \\
y_3 = [0,0,\omega^2]					& y_4 = [1,\omega,1] \\
y_5 = [0,\omega^2,0]					& y_6 = [1,1,\omega] \\
y_7 = [0,\omega^2,\omega^2]		& y_8 = [1,1,1] 
\end{array}
\end{equation*}
The $4$ groups are
\begin{equation*}
\begin{array}{cc}
\mathcal{S}_1 = \{y_1,y_7\} & \mathcal{S}_2 = \{y_2,y_8\} \\
\mathcal{S}_3 = \{y_3,y_5\} & \mathcal{S}_4 = \{y_4,y_6\}
\end{array}
\end{equation*}
It can be seen that for any two vectors in different groups the weight of their sum is odd.

\subsection{Square Orthogonal Designs} \label{SquareOD}
Square Orthogonal Designs~\cite{TiH} are square designs \mbox{${\bf X}(x_1,\dots,x_K)$} which satisfy the following equation
\begin{equation*}
{\bf X}^H{\bf X} = (\sum_{i=1}^{K}{x_i^2})I
\end{equation*}
Such designs offer both single real symbol decodability and full diversity when arbitrary real constellations are used to encode each of the real symbols $x_i$. Maximal rate square orthogonal designs were given in~\cite{TiH}. These designs are of dimension $2^m \times 2^m$ and have a rate of \mbox{$R = \frac{m+1}{2^m}$} cspcu. These designs are obtainable from codes over $\mathbb{F}_4$. There are $2m+2$ groups containing one vector each. These vectors $\{y_k\}$ are given below. For $k=1,\dots,m,$
\begin{equation*}
y_k = [\textrm{{\sf \bf 1}}\textrm{\{k is even\}},0,\dots,0,\omega^2,\omega,\dots,\omega],
\end{equation*}
\begin{equation*}
y_{k+m} = [\textrm{{\sf \bf 1}}\textrm{\{k is even\}},0,\dots,0,1,\omega,\dots,\omega],
\end{equation*}
there being $m-k$ zeros in each vector and
\begin{equation*}
y_{2m+1} = [{\sf \bf 1}\textrm{\{m is even\}},\omega,\dots,\omega],
\end{equation*}
\begin{equation*}
y_{2m+2} = [0,0,\dots,0].
\end{equation*}


\section{Known and some new Multigroup Decodable STBCs from codes over $\mathbb{F}_4$} \label{MultigroupSection}

In this section we construct multigroup decodable designs via codes over $\mathbb{F}_4$. We give three procedures to obtain a multigroup decodable design for $2^{m+1}$ antennas by using a multigroup decodable design for $2^m$ antennas. These constructions are then used to obtain $4$-group decodable codes. Lastly $g$-group decodable codes for arbitrary $g$ are constructed from codes over $\mathbb{F}_4$. These designs meet the rate-ML decoding complexity tradeoff attainable by a class of CUWDs obtainable from Extended Clifford Algebras~\cite{RaR}. 

\subsection{Construction A}
Let us denote $[1,0,\dots,0] \in \mathbb{F}_2 \oplus \mathbb{F}_4^m$ by $\delta_m$. The following proposition describes how to construct a $g$-group decodable design for $2^{m+1}$ antennas using a $g$-group decodable design for $2^m$ antennas.
\begin{proposition} \label{ABBA}
Let $l \in \{0,1,2\}$ and \mbox{$\{ \mathcal{S}_i = \{y_{i,j} \vert j = 1,\dots, \vert\Gamma_i\vert\}|i = 1,\dots,g\}$} be a $2^m \times 2^m$ $g$-group decodable design of rate $R$. Then \{$\tilde{\mathcal{S}}_i \vert i = 1,\dots,g$\} is a $2^{m+1} \times 2^{m+1}$ $g$-group decodable design of rate $R$, where $\tilde{\mathcal{S}}_i = \mathcal{S}_{i,A} \cup \mathcal{S}_{i,B}$ with
 \begin{equation*}
 \mathcal{S}_{i,A} = \{[y_{i,j},0] \vert j = 1,\dots, \vert\Gamma_i\vert\ \}
 \end{equation*}
 \begin{equation*}
 \mathcal{S}_{i,B} = \{[y_{i,j},\omega^l] + \delta_{m+1} \vert j = 1,\dots, \vert\Gamma_i\vert\ \}
 \end{equation*}
If for any $i \in \{1,\dots,g\}$
	\begin{equation*}
	{\sf wt}(y+z) \textrm{ is even for any } y,z \in \mathcal{S}_i 
	\end{equation*}
then
	\begin{equation*}
	{\sf wt}(y+z) \textrm{ is even for any } y,z \in \tilde{\mathcal{S}}_i.
	\end{equation*}	
\end{proposition}
\begin{proof}
Consider $i,j \in \{1,\dots,g\}$ and $i \neq j$. It is straightforward to show that \mbox{$\mathcal{S}_{i,A} \cap \mathcal{S}_{i,B} = \phi$} and for any $y \in \tilde{\mathcal{S}}_i$ and $z \in \tilde{\mathcal{S}}_j$, \mbox{${\sf wt}(y + z)$ is odd}. The second part of the theorem is straightforward. This completes the proof.
\end{proof}

Let \mbox{$y = [\lambda,\xi_{1},\dots,\xi_{m}] \in \mathbb{F}_2 \oplus \mathbb{F}_4^m$} and $\sigma$ be any permutation on $\{1,\dots,m\}$. Define \mbox{$\sigma : \mathbb{F}_2 \oplus \mathbb{F}_4^m \to \mathbb{F}_2 \oplus \mathbb{F}_4^m$} as
\begin{equation*}
\sigma(y) = [\lambda,\xi_{\sigma(1)},\dots,\xi_{\sigma(m)}]
\end{equation*}
$\sigma$ is thus a permutation of coordinates of $y$. In terms of linear dispersion matrices the action of $\sigma$ is to permute the order in which the $2 \times 2$ matrices appear in the Kronecker product representation. The following proposition states that $g$-group decodability of a design is not disturbed by such a permutation when it is applied to all the weight matrices.
\begin{proposition} \label{Permutation}
Let \mbox{$\mathcal{S} \subseteq \mathbb{F}_2 \oplus \mathbb{F}_4^m$} be a $g$-group decodable, FD or FGD design and $\sigma$ be any permutation on $\{1,\dots,m\}$. Then \mbox{$\tilde{\mathcal{S}}=\{\sigma(y)|y \in \mathcal{S}\}$} is a $g$-group decodable, FD or FGD design respectively. 
\end{proposition}
\begin{proof}
The action of $\sigma$ on the vectors is just a permutation of the coordinates. Thus for any \mbox{$y,z \in \mathbb{F}_2 \oplus \mathbb{F}_4^m$}, \mbox{$\sigma(y+z) = \sigma(y) + \sigma(z)$} and \mbox{${\sf wt}(\sigma(y)) = {\sf wt}(y)$}. Thus ${\sf wt}(\sigma(y) + \sigma(z)) = {\sf wt}(y + z)$. The desired result follows from Definition~\ref{LowComplexityDefinition}.
\end{proof}

Corresponding to $l=0,1$ and $2$ in Proposition~\ref{ABBA} and $\sigma$ in Proposition~\ref{Permutation} we get different constructions that give us a $g$-group decodable design for $2^{m+1}$ antennas by using a $g$-group decodable design for $2^{m}$ antennas. We now give some constructions obtained from Proposition~\ref{ABBA} as examples. Let \mbox{${\bf X}=\sum_{i=1}^{K}x_iA_i$} be a $g$-group decodable design with all the linear dispersion matrices in $\Lambda_m$ and ${\bf W}$ be an identical design in a different set of real variables. Then each of the following designs is $g$-group decodable.
\begin{equation} \label{ABBAconst}
\begin{pmatrix}
{\bf X} & {\bf W} \\
{\bf W} & {\bf X}
\end{pmatrix}
\end{equation}
\begin{equation} \label{1b}
\begin{pmatrix}
{\bf X} - {\bf W} & {\bf 0} \\
{\bf 0} & {\bf X} + {\bf W}
\end{pmatrix}
\end{equation}
\begin{equation} \label{1c}
\begin{pmatrix}
{\bf X} &  i{\bf W} \\
-i{\bf W} & {\bf X}
\end{pmatrix}
\end{equation}
Proposition~\ref{Permutation} is used along with Proposition~\ref{ABBA} to arrive at these constructions. Construction~\eqref{1b} can be used to obtain single complex symbol decodable square CIODs~\cite{KhR} by using single real symbol decodable square orthogonal designs as building blocks. Construction \eqref{ABBAconst} was first proposed in \cite{TBH} and is known as ABBA construction. As an example we describe how ABBA construction is obtained from Proposition~\ref{ABBA}.

Let  \mbox{$\{\mathcal{S}_i \vert i = 1,\dots,g\}$} be a $g$-group decodable design and $\mathcal{S} = \cup_{i=1}^{g}{\mathcal{S}_i}$. Then \mbox{$\varphi^{-1}({\mathcal{S}}) = \{A_1,\dots,A_K\}$} is the set of linear dispersion matrices. Let $\{x_1,\dots,x_K\}$ and $\{w_1,\dots,w_K\}$ be two different sets of real variables. Define \mbox{${\bf X} = \sum_{i=1}^{K}{x_iA_i}$} and \mbox{${\bf W} = \sum_{i=1}^{K}{w_iA_i}$}. Let \mbox{$\{\tilde{\mathcal{S}}_i \vert i = 1,\dots,g\}$} be the design constructed according to Proposition~\ref{ABBA} with $l=0$ and let $\tilde{\mathcal{S}} = \cup_{i=1}^{g}{\tilde{\mathcal{S}}_i}$. Then \mbox{$\varphi^{-1}({\tilde{\mathcal{S}}})$} is the set of linear dispersion matrices corresponding to the new design. From Proposition~\ref{ABBA} it is clear that \mbox{$\tilde{\mathcal{S}} = \tilde{\mathcal{S}}_A \cup \tilde{\mathcal{S}}_B$} where
\begin{equation*}
\tilde{\mathcal{S}}_A = \{[y,0] | y \in \mathcal{S}\} \mbox{   and   }
\end{equation*} 
\begin{equation*}
\tilde{\mathcal{S}}_B = \{[y,1]+\delta_{m+1}| y \in \mathcal{S}\}.
\end{equation*}
Let $\sigma$ be the permutation on $\{1,\dots,m+1\}$ given by $\sigma(1)=m+1$ and $\sigma(k)=k-1$ for $k>1$. Using Proposition~\ref{Permutation} on the design \mbox{$\{\tilde{\mathcal{S}}_i \vert i = 1,\dots,g\}$} we get a $g$-group decodable design with the set of linear dispersion matrices as \mbox{$\varphi^{-1}\circ\sigma(\tilde{\mathcal{S}}) = \varphi^{-1}\circ\sigma(\tilde{\mathcal{S}}_A) \cup \varphi^{-1}\circ\sigma(\tilde{\mathcal{S}}_B)$}. But we have
\begin{equation*}
\varphi^{-1}\circ\sigma(\tilde{\mathcal{S}}_A) = \{I_2 \otimes A_i|i=1,\dots,K\} = \left\{\begin{pmatrix} A_i & {\bf 0} \\ {\bf 0} & A_i\end{pmatrix} \right\}
\end{equation*}
\begin{equation*}
\varphi^{-1}\circ\sigma(\tilde{\mathcal{S}}_B) = \{i(iX) \otimes A_i|i=1,\dots,K\} = \left\{\begin{pmatrix} {\bf 0} & -A_i\\ -A_i & {\bf 0}\end{pmatrix} \right\}
\end{equation*}
Associating the variables $\{x_i\}$ with matrices in $\varphi^{-1}\circ\sigma(\tilde{\mathcal{S}}_A)$ and variables $\{-w_i\}$ with those in $\varphi^{-1}\circ\sigma(\tilde{\mathcal{S}}_B)$ we get the design in \eqref{ABBAconst}.

\subsection{Construction B}

The following proposition gives a procedure to obtain $2$-group decodable designs for $2^{m+1}$ antennas using $2$-group decodable designs for $2^{m}$ antennas.
\begin{proposition} \label{C1}
Let $l \in \{0,1,2\}$ and  \mbox{$\{ \mathcal{S}_i = \{y_{i,j} \vert j = 1,\dots, \vert\Gamma_i\vert\}|i = 1,2\}$} be a \mbox{$2^m \times 2^m$} $2$-group decodable design which satisfies the following condition for each $i =1,2$
	\begin{equation*}
	{\sf wt}(y+z) \textrm{ is even for any } y,z \in \mathcal{S}_i 
	\end{equation*}
then $\tilde{\mathcal{S}}_1 = \mathcal{S}_{1,A} \cup \mathcal{S}_{2,B}$, $\tilde{\mathcal{S}}_2 = \mathcal{S}_{2,A} \cup \mathcal{S}_{1,B}$ give a $2^{m+1} \times 2^{m+1}$ $2$-group decodable design \mbox{$\{\tilde{\mathcal{S}}_1,\tilde{\mathcal{S}}_2\}$} which satisfies for each $i = 1,2$
	\begin{equation*}
	{\sf wt}(y+z) \textrm{ is even for any } y,z \in \tilde{\mathcal{S}}_i 
	\end{equation*}
where
	\begin{equation*}
	\mathcal{S}_{i,A} = \{[y_{i,j},0]\} \textrm{ and } \mathcal{S}_{i,B} = \{[y_{i,j},\omega^l]\}
	\end{equation*}
Further the rates of $\{\mathcal{S}_i\}$ and $\{\tilde{\mathcal{S}}_i\}$ are same.
\end{proposition} 
\begin{proof}
Similar to the proof of Proposition~\ref{ABBA}.
\end{proof}

As particular examples we get the following recursive constructions. Let ${\bf X}$ and ${\bf W}$ be identical $2$-group decodable designs in different variables with the weight matrices in $\Lambda_m$. Let the design ${\bf X}$ be such that the corresponding vectors in \mbox{$\mathbb{F}_2 \oplus \mathbb{F}_4^m$} satisfy the hypothesis of Proposition~\ref{C1}. The following designs are $2$-group decodable.
\begin{equation} \label{1d}
\begin{pmatrix}
{\bf X} & i{\bf W} \\
i{\bf W} & {\bf X}
\end{pmatrix}
\end{equation}
\begin{equation} \label{1e}
\begin{pmatrix}
{\bf X} + i{\bf W} & {\bf 0} \\
{\bf 0} & {\bf X} - i{\bf W}
\end{pmatrix}
\end{equation}
\begin{equation} \label{1f}
\begin{pmatrix}
{\bf X} &  {\bf W} \\
-{\bf W} & {\bf X}
\end{pmatrix}
\end{equation} 
These designs can be obtained from Proposition~\ref{C1} in the same way the ABBA construction was obtained from Proposition~\ref{ABBA}.

\subsection{Construction C}
The following proposition gives a procedure to obtain $4$-group decodable designs for $2^{m+1}$ antennas using $2$-group decodable designs for $2^{m}$ antennas.
\begin{proposition} \label{C2}
Let \mbox{$\{ \mathcal{S}_i = \{y_{i,j} \vert j = 1,\dots, \vert\Gamma_i\vert\}|i = 1,2\}$} be a $2^m \times 2^m$ $2$-group decodable design which satisfies for each $i=1,2$
	\begin{equation*}
	{\sf wt}(y+z) \textrm{ is even for any } y,z \in \mathcal{S}_i 
	\end{equation*}
then $\{\tilde{\mathcal{S}}_i|i = 1,\dots,4\}$ is a $2^{m+1} \times 2^{m+1}$ $4$-group decodable design where
	\begin{equation*}
	\tilde{\mathcal{S}}_1 = \{[y_{1,j},\xi_1]\} \textrm{,  } \tilde{\mathcal{S}}_2 = \{[y_{1,j},\xi_2]\}
	\end{equation*}
	\begin{equation*}
	 \tilde{\mathcal{S}}_3 = \{[y_{2,j},\xi_3] + \delta_{m+1}\} \textrm{,  } \tilde{\mathcal{S}}_4 = \{[y_{2,j},\xi_4] + \delta_{m+1}\} 
	\end{equation*}
and $\xi_i$, $i=1,\dots,4$ are distinct elements of $\mathbb{F}_4$. Further the rates of $\{\mathcal{S}_i\}$ and $\{\tilde{\mathcal{S}}_i\}$ are same.
\end{proposition}
\begin{proof}
Similar to the proof of Proposition~\ref{ABBA}.
\end{proof}

There are $4! = 24$ ways of choosing $\xi_i,i=1,\dots,4$ from $\mathbb{F}_4$. However, it can be shown that \mbox{$\{\xi_1,\xi_2,\xi_3,\xi_4\}$}, \mbox{$\{\xi_1,\xi_2,\xi_4,\xi_3\}$}, \mbox{$\{\xi_2,\xi_1,\xi_3,\xi_4\}$} and \mbox{$\{\xi_2,\xi_1,\xi_4,\xi_3\}$} all lead to designs which are same upto relabeling of variables. Thus Proposition~\ref{C2} gives us $6$ constructions. However only $4$ of them are unique i.e. lead to non-equivalent designs. Two others can be obtained by permutation of columns and relabeling of variables of one of the $4$ designs. These $4$ constructions correspond to $\{0,1,\omega,\omega^2\}$, $\{\omega,\omega^2,0,1\}$, $\{1,\omega^2,0,\omega\}$ and $\{\omega,1,0,\omega^2\}$.

\subsection{$4$-group decodable designs}

We now give a procedure to get a $4$-group decodable design for $2^m$ antennas, $m \geq 1$, using any $2$-group decodable design \mbox{$\{\mathcal{S}_1,\mathcal{S}_2\}$} for $2^{m-k}$ antennas, $k \geq 1$, which satisfies the following condition for each $i=1,2$
\begin{equation}
{\sf wt}(y+z) \textrm{ is even for any } y,z \in  \mathcal{S}_i
\end{equation}
Define Step A and Step B as the following steps.
\begin{itemize}

\item Step A: Apply any one of the $6$ constructions choosing from Propositions~\ref{ABBA} and~\ref{C1} and $l=0,1$ or $2$. Follow it by an application of Proposition~\ref{Permutation} with any $\sigma$.

\item Step B: Apply any one of the $4$ constructions choosing from those provided by Proposition~\ref{C2}. Follow it by an application of Proposition~\ref{Permutation} with any $\sigma$.

\end{itemize}
The construction procedure is as follows: Starting with the design \mbox{$\{\mathcal{S}_1,\mathcal{S}_2\}$} apply Step A $k-1$ times followed by one application of Step B.

Particular examples of this procedure are the constructions given in \cite{Abr} and \cite{RTR}. We now explain how these constructions are particular applications of the above algorithm. To explain this we need the following proposition.
\begin{proposition} \label{EvenOdd}
Let \mbox{$\{\tilde{\mathcal{S}}_1,\tilde{\mathcal{S}}_2\}$} be a $2$-group decodable design obtained through the application of any of the constructions of Propositions~\ref{ABBA} or~\ref{C1} on the $2$-group decodable design \mbox{$\{{\mathcal{S}}_1,{\mathcal{S}}_2\}$}. If ${\mathcal{S}}_1,{\mathcal{S}}_2$ have even and odd weight vectors respectively, then $\tilde{\mathcal{S}}_1,\tilde{\mathcal{S}}_2$ have even and odd weight vectors.
\end{proposition}
\begin{proof}
Straightforward.
\end{proof}
Let \mbox{$\mathcal{M}_i = \varphi^{-1}(\mathcal{S}_i)$} be the $i^{th}$ group of linear dispersion matrices. Both constructions,\cite{Abr} and \cite{RTR}, start with the trivial design for one antenna, ${\bf X} = (x_1 + ix_2)$. This design satisfies the hypothesis of Proposition~\ref{EvenOdd}. Thus, at the end of $k-1$ applications of Step A, the resulting code $\{\mathcal{S}_1,\mathcal{S}_2\}$ will be such that $\mathcal{M}_1$ has Hermitian and $\mathcal{M}_2$ has skew-Hermitian matrices. In such a scenario the matrix representation of the four constructions in Proposition~\ref{C2} are given as follows. 

Let \mbox{$\{\mathcal{S}_1,\mathcal{S}_2\}$} be a two group decodable design satisfying the hypotheses of Propositions~\ref{C2} and~\ref{EvenOdd}. Let ${\bf X}$ be the design obtained from \mbox{$\{\mathcal{S}_1,\mathcal{S}_2\}$} and let ${\bf W}$ be identical to ${\bf X}$  but be composed of a different set of variables. Define for any square matrix $A$,  \mbox{$A_H = \frac{1}{2}(A + A^H)$} and \mbox{$A_{SH} = \frac{1}{2}(A - A^H)$}. These are the Hermitian and skew-Hermitian parts of $A$. The following $4$-group decodable designs can be obtained from Proposition~\ref{C2}.
\begin{equation} \label{2a}
\begin{pmatrix}
{\bf X^H} & i{\bf W} \\
i{\bf W^H} & {\bf X}
\end{pmatrix}
\end{equation}
\begin{equation} \label{2b}
\begin{pmatrix}
i{\bf X} & {\bf W^H} \\
-{\bf W} & -i{\bf X^H}
\end{pmatrix}
\end{equation}
\begin{equation} \label{2c}
\begin{pmatrix}
i{\bf X_{SH}}-{\bf W_{SH}} & {\bf W_{H}}+i{\bf X_{H}} \\
-{\bf W_{H}}+i{\bf X_{H}} & i{\bf X_{SH}}+ {\bf W_{SH}}
\end{pmatrix}
\end{equation}
\begin{equation} \label{2d}
\begin{pmatrix}
{\bf X} & {\bf W} \\
-{\bf W^H} & {\bf X^H}
\end{pmatrix}
\end{equation}
The above constructions can be obtained in a way similar to which ABBA construction was obtained from Proposition~\ref{ABBA} and by using the fact that $\mathcal{M}_1$ has Hermitian and $\mathcal{M}_2$ has skew-Hermitian matrices. 

Constructions in \cite{Abr}  and \cite{RTR} start with ${\bf X} = (x_1 + ix_2)$. Constructions in  \cite{RTR} use either \eqref{ABBAconst} or \eqref{1f} for the first application of Step A and uses \eqref{ABBAconst} for each of the remaining $k-2$ applications of Step A. The last step in \cite{RTR} is the application of
\begin{equation} \label{DoublingConstruction}
\begin{pmatrix}
{\bf X} & -{\bf W^H} \\
{\bf W} & {\bf X^H}
\end{pmatrix}
\end{equation}
for Step B. This construction was first given in~\cite{KiR} and is known as the Doubling Construction. But this is same as~\eqref{2d} upto relabeling of variables. Constructions in ~\cite{Abr} use \eqref{1f} for each of the $k-1$ applications of Step A and \eqref{2d} for Step B.

\subsection{$g$-group decodable designs for arbitrary $g$}
In this section we construct $g$-group decodable designs with arbitrary $g$ for the case when the number of real symbols in each group is same and is equal to a power of two i.e. \mbox{$\tau = 2^a$}. Resulting designs are for number of transmit antennas $2^{b}$, where
\begin{equation*}
b \geq \left\lceil \frac{g}{2} - 1 \right\rceil 
\end{equation*}

Let us first consider the case where $g$ is even. Say \mbox{$g = 2(m+1)$}. We start with a square orthogonal design for $2^m$ antennas. We already saw that square orthogonal designs are obtainable from $\mathbb{F}_2 \oplus \mathbb{F}_4^m$. Such a design has rate $R=\frac{m+1}{2^m}$ and has $2m+2$ groups with one real symbol per group. Now we apply Proposition~\ref{ABBA} on this design $a$ times. Each of the applications can use any of the three constructions given in Proposition~\ref{ABBA} and can be followed with an application of Proposition~\ref{Permutation} with arbitrary $\sigma$. 

According to Propositions~\ref{ABBA} and~\ref{Permutation} the resulting code will be for $2^{m+a}$ antennas, with $g = 2m+2$ groups and rate $R=\frac{m+1}{2^m}$. Number of real symbols will be
\begin{equation*}
K = 2 \times R \times \textrm{ Number of antennas }= 2(m+1)2^a.
\end{equation*}
Therefore, the number of real symbols per group $\tau = 2^a$ as required. The rate in terms of $g$ is \mbox{$R = \frac{g}{2^{g/2}}$}.

Now consider the case when $g$ is odd. Suppose $g=2m+1$ for some $m$, define $g' = g+1 = 2m+2$. Since $g'$ is even we can construct a $g'$-group decodable design for $\tau = 2^a$ as described above. This design for $2^{m+a}$ antennas will have $g+1$ groups. This is more than what is required. The desired design is obtained by removing any one group from this design. The rate of the resulting design is 
\begin{equation*}
R = \frac{1}{2}\frac{\tau g}{2^{m+a}} = \frac{g}{2^{\frac{g+1}{2}}}.
\end{equation*}

Thus, for an arbitrary $g,$ a rate of
\begin{equation} \label{RateEqn}
R=\frac{g}{2^{\lfloor \frac{g+1}{2} \rfloor}}
\end{equation}
is achievable. Since a $g$-group decodable square orthogonal design exists only for $2^{\lceil \frac{g}{2} - 1 \rceil}$ or more antennas, the construction procedure described above can be used to get $g$-group decodable designs for any number of transmit antennas $2^b$ with \mbox{$b \geq \lceil \frac{g}{2} - 1 \rceil $}. In \cite{RaR} the rate-ML decoding complexity tradeoff of the class of CUWDs for which $\tau$ is a power of $2$ was characterized. The maximum rate, as given in \cite{RaR}, of any CUWD for a given $g$ and $\tau=2^a$ is precisely \eqref{RateEqn}.


\renewcommand{\arraystretch}{1.25}
\begin{center}
\begin{table*}[htbp]

\caption{Comparison of Decoding Complexities: A - Arbitrary constellation,    B - Carefully chosen constellation   $\ast$  - Least known complexity}
\begin{tabular} {|l|l|l|l|l|l|l|l|}
\hline 
\multirow{2}{*}{Transmit Antennas $N$} & \multirow{2}{*}{Rate $R$}	& \multicolumn{2}{c|}{New codes in Sec~\ref{NewClass}} & EAST Codes & Pavan et al.~\cite{SrR2} & New codes in Sec~\ref{FromPavan} & FGD Code  \\ 
\cline{3-4} \cline{6-7}
                                 &          & A             & B              &    Sinnokrot et al.~\cite{SBM}        &    A    &   B                   &  Ren et al.~\cite{RGYS}   \\ 
\hline 
$2$ & $2$ & $\ast2M^{3}$ & $\ast3M^2$ &   &   &   &   \\ 
\hline
\multirow{5}{*}{$4$} & $5/4$ & $\ast3M^{2}$ & $\ast3M^{1.5}$ &  & $2M^{2.5}$ & $2M^2$ &  \\
\cline{2-8}
& $2$ & $\ast3M^5$ & $\ast3M^{4.5}$ & $4M^5$ & & & $5M^{5.5}$ \\ 
\cline{2-8}
& $17/8$ & $\ast3M^{5.5}$ & $\ast3M^5$ & & & & $5M^6$ \\
\cline{2-8}
& $3$ & $\ast3M^{9}$ & $\ast3M^{8.5}$ & & & & \\
\cline{2-8}
& $4$ & $\ast3M^{13}$ & $\ast3M^{12.5}$ & & & & \\
\hline                                       
\multirow{7}{*}{$8$} & $5/4$ & $\ast3M^4$ & $\ast3M^{3.5}$ & & $2M^5$ & $2M^{4.5}$ & \\
\cline{2-8}
& $2$ & $3M^{10}$ & $3M^{9.5}$ & $4M^{10}$ & $\ast2M^8$ & $\ast2M^{7.5}$ & \\
\cline{2-8}
& $17/8$ & $3M^{11}$ & $3M^{10.5}$ & & $\ast2M^{8.5}$ & $\ast2M^8$ & \\
\cline{2-8}
& $3$ & $\ast3M^{18}$ & $\ast3M^{17.5}$ & $4M^{18}$ & & & \\
\cline{2-8}
& $4$ & $\ast3M^{26}$ & $\ast3M^{25.5}$ & $4M^{26}$ & & & \\
\cline{2-8}
& $5$ & $\ast3M^{34}$ & $\ast3M^{33.5}$ & & & & \\
\cline{2-8}
& $6$ & $\ast3M^{42}$ & $\ast3M^{42.5}$ & & & & \\
\hline
\end{tabular}
\label{Comparison}
\end{table*}
\end{center}
\renewcommand{\arraystretch}{1}


\section{New Fast-Group-Decodable Codes} \label{FDSection}
In this section we propose a new class of fast-group-decodable and fast-decodable codes with rates $R > 1$ for number of antennas that are a power of $2$. The rate-decoding complexity tradeoff of this class of codes is derived. We also show that codes with lower decoding complexity than those reported in~\cite{SrR2} can be obtained by simply using the same design as in~\cite{SrR2} but by choosing the constellations carefully. We also show that the STBC given in~\cite{RGYS} is a specific case of STBCs obtained from codes over $\mathbb{F}_4$. In the latter part of this section we show that some of the best known codes for $2$ and $4$ transmit antennas are obtainable from codes over $\mathbb{F}_4$. 

\subsection{A new class of FD and FGD designs} \label{NewClass}
We first propose a new class of rate $5/4$ fast-group-decodable designs. These designs are then extended to obtain fast-decodable designs with rates $R>5/4$. Designs of rate less than $5/4$ are obtained by puncturing.

Let the number of transmit antennas be $2^m$, $m \geq 1$. Let \mbox{$\xi_1,\xi_2 \in \mathbb{F}_4 \setminus \{0\}$} and \mbox{$\xi_1 \neq \xi_2$}. Let $\xi_3=\xi_1 + \xi_2$. Define 
\begin{equation*}
\mathcal{S}_{\xi_1}= \{[0,\zeta_1,\dots,\zeta_m] | \zeta_i \in \{0,\xi_1\} \textrm{ for } i=1,\dots,m\},
\end{equation*}
\begin{equation*}
\mathcal{S}_A = \{ y \in \mathcal{S}_{\xi_1} | {\sf wt}(y) \textrm{ is even } \} \mbox{  and  }
\end{equation*}
\begin{equation*}
\mathcal{S}_B = \{ y \in \mathcal{S}_{\xi_1} | {\sf wt}(y) \textrm{ is odd } \}.
\end{equation*}
Let \mbox{$\nu_m = [{\sf \bf 1}\{\textrm{m is even}\},\xi_2,\dots,\xi_2]$} and \mbox{$\delta_m = [1,0,\dots,0]$}. Define
\begin{equation*}
\mathcal{S}_C = \nu_m + \mathcal{S}_A, ~~
\mathcal{S}_D = \nu_m + \mathcal{S}_B, ~~ \mbox{ and } 
\mathcal{S}_E = \delta_m + \mathcal{S}_A.
\end{equation*}

Let $\mathcal{S}_1 = \mathcal{S}_A$ and \mbox{$\mathcal{S}_2 = \cup_{j \in \{B,C,D,E\}}{\mathcal{S}_j}$}. Then we have the following propositions.
\begin{proposition} \label{FDprop1}
$\{\mathcal{S}_A,\mathcal{S}_B,\mathcal{S}_C,\mathcal{S}_D\}$ is a $4$-group decodable, rate $1$ design. 
\end{proposition}
\begin{proof}
It must be noted that $\mathcal{S}_A$ is a subgroup of the abelian group \mbox{$\mathbb{F}_2 \oplus \mathbb{F}_4^m$} and \mbox{$\mathcal{S}_B = \gamma_m + \mathcal{S}_A$}, where \mbox{$\gamma_m = [0,0,\dots,0,\xi_1]$}. Thus $\mathcal{S}_B$, $\mathcal{S}_C$ and $\mathcal{S}_D$ are cosets of the subgroup $\mathcal{S}_A$ and are obtained by the translates $\gamma_m$, $\nu_m$ and \mbox{$\gamma_m + \nu_m$} respectively. It is straightforward to see that all three cosets have only odd weight vectors. Also $\{0,\gamma_m,\nu_m,\gamma_m+\nu_m\}$ is a subgroup of \mbox{$\mathbb{F}_2 \oplus \mathbb{F}_4^m$}. Because both $\mathbb{F}_2$ and $\mathbb{F}_4$ have characteristic $2$, every element of \mbox{$\mathbb{F}_2 \oplus \mathbb{F}_4^m$} is its own inverse.

Let $i,j \in \{A,B,C,D\}$ and $i \neq j$. Let $y_i \in \mathcal{S}_i$ and $y_j \in \mathcal{S}_j$. Then there exist \mbox{$u_i,u_j \in \mathcal{S}$} and \mbox{$w_i,w_j \in \{0,\gamma_m,\nu_m,\gamma_m+\nu_m\}$} with \mbox{$w_i \neq w_j$} such that \mbox{$y_i = w_i + u_i$} and \mbox{$y_j = w_j + u_j$}. Thus \mbox{$y_i + y_j = w_i + w_j + u_i + u_j = w + u$} for some \mbox{$u \in \mathcal{S}_A$} and \mbox{$w \in \{\gamma_m,\nu_m,\gamma_m+\nu_m\}$}. Thus \mbox{$y_i + y_j$} is an element of \mbox{$\cup_{k \in \{B,C,D\}}{\mathcal{S}_k}$} and hence has odd weight. From Definition~\ref{LowComplexityDefinition} the given design is $4$-group decodable.

The number of elements in $\mathcal{S}_{\xi_1}$ is $2^m$. Thus \mbox{$|\mathcal{S}_j|=2^{m-1}$} for \mbox{$j \in \{A,B,C,D\}$}. It is straightforward to show that the four subsets are mutually non-intersecting. Thus the rate of the proposed design is $1$ complex symbol per channel use. This completes the proof.
\end{proof}

\begin{proposition} \label{FDprop2}
The design \mbox{$\{\mathcal{S}_1,\mathcal{S}_2\}$} is $2$-group decodable.
\end{proposition}
\begin{proof}
Since \mbox{$\{\mathcal{S}_A,\mathcal{S}_B,\mathcal{S}_C,\mathcal{S}_D\}$} was already shown to be $4$-group decodable in Proposition~\ref{FDprop1}, it is enough to show that for every \mbox{$y_1 \in \mathcal{S}_A$} and \mbox{$y_2 \in \mathcal{S}_E$}, \mbox{$y_1 + y_2$} has odd weight. Now $\mathcal{S}_E$ is a coset of the additive subgroup $\mathcal{S}_A$ and hence \mbox{$y_1 + y_2 \in \mathcal{S}_E$}. But every vector in $\mathcal{S}_E$ has odd weight. This completes the proof.
\end{proof}


From Propositions~\ref{FDprop1} and~\ref{FDprop2} we see that \mbox{$\{\mathcal{S}_1,\mathcal{S}_2\}$} is a fast-group-decodable design of rate $5/4$. The design $\mathcal{S}_2$ is conditionally $3$-group decodable with the conditional groups as $\mathcal{S}_B$, $\mathcal{S}_C$ and $\mathcal{S}_D$. For rates $R \geq 5/4,$ choose any subset \mbox{$\mathcal{O} \subseteq \mathbb{F}_2 \oplus \mathbb{F}_4^m \setminus \{\mathcal{S}_1 \cup \mathcal{S}_2 \}$} with \mbox{$|\mathcal{O}| = 2^{m-1}(4R-5)$}. This can always be done as long as $R$ is less than or equal to the maximum possible rate $2^m$. The proposed design is
\begin{equation}
\mathcal{S} = \mathcal{S}_1 \cup \mathcal{S}_2 \cup \mathcal{O} \label{NewFD}.
\end{equation}

It is straightforward to show that the design~\eqref{NewFD} has rate $R$. We now derive its ML decoding complexity for arbitrary complex constellations for the case when each $\mathcal{S}_j$, \mbox{$j \in \{A,B,C,D\}$} has integral number of complex symbols. This happens whenever $m>1$. The complexity of decoding $\mathcal{S}$ is 
\begin{equation*}
M^{\frac{1}{2}|\mathcal{O}|} \times \textrm{ decoding complexity of } \mathcal{S}_1 \cup \mathcal{S}_2.
\end{equation*}
Observing that \mbox{$\{\mathcal{S}_1,\mathcal{S}_2\}$} is $2$-group decodable and $\mathcal{S}_2$ is conditionally $3$-group decodable, we see that the decoding complexity of \mbox{$\mathcal{S}_1 \cup \mathcal{S}_2$} is
\begin{equation*}
M^{\frac{1}{2}2^{m-1}} + 3M^{2^{m-1}}.
\end{equation*}
Thus the decoding complexity of the proposed design for arbitrary complex constellations is 
\begin{align}
&= M^{2^{m-2}(4R-5)} \times ( M^{\frac{1}{2}2^{m-1}} + 3M^{2^{m-1}} ) \nonumber \\
&\simeq 3M^{2^{m-1}+2^{m-2}(4R-5)} \nonumber \\
&= 3M^{2^{m-2}(4R-3)} \textrm{ for } m>1 \textrm{, } R \geq \frac{5}{4} \label{Rate-Complexity1}.
\end{align}

The ML decoding complexity can be further reduced by using carefully chosen constellations without affecting full diversity. We now show how to choose constellations for this purpose and derive the ML decoding complexity for this case. Assume that $m\geq1$ and $R\geq5/4$. Let $x_1$,$x_2$,$x_3$ and $x_4$ be four real symbols one each from $\mathcal{S}_A$, $\mathcal{S}_B$, $\mathcal{S}_C$ and $\mathcal{S}_D$ respectively. Since the corresponding weight matrices $A_i$ are unitary, they satisfy
\begin{equation*}
A_i^HA_j + A_j^HA_i = {\bf 0} \textrm{ for } 1\leq i < j \leq 4.
\end{equation*} 
By using Theorem~\ref{RegularPAMTheorem} we see that each of the symbols $x_i$, $i=1,2,3,4$ can be assigned values from a regular PAM. During decoding, when values are proposed for all real symbols except $x_i$, $i=1,2,3,4$, these four symbols can be decoded via simple scaling and rounding off. In this case the decoding complexity of the code is 
\begin{equation}
3M^{2^{m-2}(4R-3)-0.5} \textrm{ for } m\geq1 \textrm{, } R \geq \frac{5}{4}\label{Rate-Complexity2}.
\end{equation}
Similarly it can be shown that for rates \mbox{$1\leq R \leq 5/4$} the designs obtained by puncturing the vectors in $\mathcal{S}_E$ give rise to decoding complexities which have the same expression as in~\eqref{Rate-Complexity1} and~\eqref{Rate-Complexity2}. 

Thus, the rate-ML decoding complexity tradeoff of the given class of codes for $R > 1$ is
\begin{equation}
3M^{2^{m-2}(4R-3)}, ~~  m>1  \label{Rate-Complexity11} \\ 
\end{equation}
for arbitrary complex constellations and 
\begin{align}
3M^{2^{m-2}(4R-3)-0.5}, ~~ m \geq 1  \label{Rate-Complexity22}
\end{align}
for appropriately chosen constellations.

\subsection{Complexity reducing constellations for designs in~\cite{SrR2}} \label{FromPavan}
The codes in~\cite{SrR2} are $2$-group decodable with unitary weight matrices and rate \mbox{$R=2^{m-2}+\frac{1}{2^m}$}. In~\cite{SrR2} decoding complexity was given only for arbitrary complex constellations, which is
\begin{equation} \label{PavanHighRateComplexity1}
2M^{2^{m-1}R}.
\end{equation}
Consider any two symbols, $x_1$ and $x_2$, one from each of the two groups. Since the linear dispersion matrices are unitary, the weight matrices $A_1$, $A_2$ of the symbols $x_1$ and $x_2$ satisfy
\begin{equation*}
A_1^HA_2 + A_2^HA_1 = {\bf 0}.
\end{equation*}
We can use Theorem~\ref{RegularPAMTheorem} to use regular PAM on two variables $x_1$ and $x_2$ without losing full diversity property. Decoding complexity now becomes
\begin{equation} \label{PavanHighRateComplexity2}
2M^{2^{m-1}R-0.5}.
\end{equation}

\subsection{Code in~\cite{RGYS} as a specific case of STBCs via codes over $\mathbb{F}_4$}
It was shown in Section~\ref{SquareOD} that square orthogonal designs belong to the class of codes obtainable from codes over $\mathbb{F}_4$. Consider the case of $m=2$. A square orthogonal design for $2^2$ antennas has $6$ vectors each forming a group on its own. One of the vectors is the all zero vector. Thus the remaining $5$ vectors are of odd weight. Let $\mathcal{O}$ be the set of these $5$ vectors. Consider the following $2$-group decodable design $\{\mathcal{S}_1,\mathcal{S}_2\}$ where
\begin{equation} 
\mathcal{S}_1 = \{[0,\dots,0]\} \textrm{, } \mathcal{S}_2 = \{y \in \mathbb{F}_2 \oplus \mathbb{F}_4^2| {\sf wt}(y) \textrm{ is odd}\}. \label{FGD1}
\end{equation}
Thus $\mathcal{O} \subseteq \mathcal{S}_2$. Further $\mathcal{O}$, when considered as a design by itself, is single real symbol decodable or $5$-group decodable. Thus the design in \eqref{FGD1} is fast-group-decodable. Since $16$ vectors are of odd weight of the total of $32$ vectors in \mbox{$\mathbb{F}_2 \oplus \mathbb{F}_4^2$}, $|\mathcal{S}_2|=16$. Hence, the above design has a rate of $17/8$ complex symbols per channel use. 

The decoding complexity of the code \eqref{FGD1} is the sum of the decoding complexities of $\mathcal{S}_1$ and $\mathcal{S}_2$. The decoding complexity of $\mathcal{S}_1$ is $M^{\frac{1}{2}}$. When decoding $\mathcal{S}_2$, for each set of values assigned to the real variables corresponding to $\mathcal{S}_2\setminus\mathcal{O}$, the real variables corresponding to $\mathcal{O}$ can be conditionally decoded with a complexity of $5M^{\frac{1}{2}}$. The net complexity of decoding $\mathcal{S}_2$ would be the product of this term with $M^{\frac{1}{2}(|\mathcal{S}_2\setminus\mathcal{O}|)}$, which is $ 5M^{\frac{1}{2}} \times M^{\frac{1}{2}(2^{4}-5)} = 5M^{\frac{1}{2}(17-5)} = 5M^{6}.$ Thus the complexity of decoding the code \eqref{FGD1} is $5M^{6} + M^{\frac{1}{2}} \simeq 5M^{6}.$ This design was the one proposed in~\cite{RGYS}.

\subsection{Comparison of decoding complexities}
From~\eqref{Rate-Complexity11},~\eqref{Rate-Complexity22},~\eqref{PavanHighRateComplexity1} and~\eqref{PavanHighRateComplexity2} it is clear that the new class of designs described in Section~\ref{NewClass} yield lower decoding complexity than the codes in~\cite{SrR2} whenever $R < 3/2$ or \mbox{$R > 2^{m-2}+\frac{1}{2^m}$}. In fact these codes have the least known complexity for these values of $R$ when the number of antennas is $2^m$. For rates \mbox{$3/2 \leq R \leq 2^{m-2}+\frac{1}{2^m}$} the proposed codes in Section~\ref{FromPavan} have the least known complexity. Table~\ref{Comparison} summarizes the comparison of the decoding complexity of known codes and the new ones of this paper. Only rates higher than $1$ are considered. Comparison is done with EAST(Embedded Alamouti Space-Time) codes from~\cite{SBM}, $2$-group decodable codes from~\cite{SrR2} and the FGD code from~\cite{RGYS}. The entry for $2$ antennas with rate $2$ and arbitrary constellation is that of the code given in~\cite{HTW} and~\cite{PGA}. In Section~\ref{FurtherExamplesHTWPGA}, it is shown that this code belongs to the proposed class of STBCs. It must be noted that the proposed code for $N=4$, $R=5/4$ has lower decoding complexity than the code from~\cite{YGT1},~\cite{SrR2}. The code for $N=4$, $R=17/8$ has lower decoding complexity than the code from~\cite{RGYS}. Similarly, for $N=8$ and $R=5/4$ the proposed codes have the least complexity.

\subsection{Examples of FD codes in literature obtainable from codes over $\mathbb{F}_4$} 

\subsubsection{Rate $2$ Codes from Pavan et al.~\cite{SrR1}}
In~\cite{SrR1} rate $2$ codes were given for $2$ and $4$ transmit antennas with the largest known coding gain. Both these codes are fast-decodable. These codes can be obtained from Pauli weight matrices by using appropriate signal sets. We emphasize that these codes have low ML decoding complexity because the underlying design comes from $\Lambda_m$. As an example we now show how the \mbox{$2 \times 2$} code is obtained from $\Lambda_1$.

Choose weight matrices from $\Lambda_1$ as 
\begin{equation*}
\begin{array}{ll}
A_1 = I_2,  & A_5 = X, \\
A_2 = Z,  & A_6 = ZX, \\
A_3 = iI_2,  & A_7 = iX, \\
A_4 = iZ,  & A_8 = iZX. \\
\end{array}
\end{equation*}
The corresponding vectors are
\begin{equation*}
\begin{array}{ll}
y_1 = [0,0],  & y_5 = [1,1], \\
y_2 = [1,\omega],  & y_6 = [0,\omega^2], \\
y_3 = [1,0],  & y_7 = [0,1], \\
y_4 = [0,\omega],  & y_8 = [1,\omega^2]. \\
\end{array}
\end{equation*}
The resulting design \mbox{${\bf X} = \sum_{i=1}^{8}{x_iA_i}$} is
\begin{equation*} \label{PavanLSTD}
\begin{pmatrix}
(x_1 + x_2) + i(x_3+x_4) & (x_5 + x_6) + i(x_7+x_8) \\
(x_5 - x_6) + i(x_7-x_8) & (x_1 - x_2) + i(x_3-x_4)   
\end{pmatrix}.
\end{equation*}
Note that the rate $1$ design \mbox{$\{y_1,y_1,y_3,y_4\}$} is two group decodable with the two groups being \mbox{$\{y_1,y_2\}$} and \mbox{$\{y_3,y_4\}$}. When the symbols in the design ${\bf X}$ are encoded in $3$-groups \mbox{$\{x_1,x_2\}$}, \mbox{$\{x_3,x_4\}$} and \mbox{$\{x_5,x_6,x_7,x_8\}$} we see that the resulting STBC is conditionally $2$-group decodable, the two groups being \mbox{$\{x_1,x_2\}$} and \mbox{$\{x_3,x_4\}$}. This leads to low complexity ML decoding.

In~\cite{SrR1} the three groups are encoded as follows. Let \mbox{$s_k = s_{k,I}+is_{k,Q}$}, \mbox{$k=1,2,3,4$} take values independently from a rotated QAM constellation. The angle of rotation is optimized for coding gain. Encode $x_i$, $i=1,\dots,8$ as follows
\begin{equation*}
\begin{pmatrix} x_1 \\ x_2 \end{pmatrix} = \frac{1}{2}\begin{pmatrix} 1 & -1 \\ 1 & 1 \end{pmatrix} \begin{pmatrix} s_{1,I} \\ s_{1,Q} \end{pmatrix},
\end{equation*}
\begin{equation*}
\begin{pmatrix} x_3 \\ x_4 \end{pmatrix} = \frac{1}{2}\begin{pmatrix} 1 & 1 \\ -1 & 1 \end{pmatrix} \begin{pmatrix} s_{2,I} \\ s_{2,Q} \end{pmatrix} \mbox{  and  },
\end{equation*}
\begin{equation*}
\begin{pmatrix} x_5\\x_6\\x_7\\x_8 \end{pmatrix} = \frac{1}{2\sqrt{2}} \begin{pmatrix} 1 & -1 & 1 & 1 \\
								1 & -1 & -1 & -1 \\
																																											 1 & 1 & 1 & -1 \\
																																											 1 & 1 &-1 & 1
																																											 \end{pmatrix} \begin{pmatrix} s_{4,I}\\s_{3,Q}\\s_{3_I}\\s_{4,Q}\end{pmatrix}
\end{equation*}
The resulting design in terms of $\{s_k\}$ is
\begin{equation} \label{PavanCode}
\begin{pmatrix}
s_{1,I}+is_{2,Q} & e^{i\pi/4}(s_{4,I}+is_{3,Q}) \\
e^{i\pi/4}(-s_{4,Q}+is_{3,I}) & -s_{1,Q}+is_{2,I}
\end{pmatrix}
\end{equation}
The STBC presented in~\cite{SrR1} is~\eqref{PavanCode} multiplied on the right hand side by the unitary matrix $\begin{pmatrix} 1 & 0 \\ 0 & -i\end{pmatrix}$.

\subsubsection{The HTW-PGA Code} \label{FurtherExamplesHTWPGA}
This is a rate $2$ code for $2$ transmit antennas. It was was independently discovered by Hottinen, Tirkkonen and Wichman~\cite{HTW} and by Paredes, Gershman and Alkhansari~\cite{PGA}. Its decoding complexity is of the order of $M^3$ for arbitrary constellations and $M^2$ for QAM symbols~\cite{SrR1}. We now show that this code is obtained from an design with Pauli Weight matrices. Specifically the rate $2$ design for $2$ transmit antennas described in Section~\ref{NewClass} leads to this code.

The HTW-PGA code in complex symbols \mbox{$s_1,s_2,s_3,s_4$} is 
\begin{equation} \label{HTWPGA1}
{\bf X} = \begin{pmatrix}
					s_1 & s_2 \\
					-\bar{s}_2 & \bar{s}_1
					\end{pmatrix}	+ \begin{pmatrix}
														s_3 & s_4 \\
													-\bar{s}_4 & \bar{s}_3
					\end{pmatrix}						\begin{pmatrix}
																		1 & 0 \\
																		0 & -1
																	\end{pmatrix},
\end{equation}
where $s_1,s_2$ are independent complex symbols and \mbox{$s_3,s_4$} are obtained from independent complex symbols $z_3,z_4$ via a unitary matrix $U$
\begin{equation} \label{HTWPGA2}
\begin{pmatrix}
s_3 \\
s_4
\end{pmatrix} = U \begin{pmatrix} z_3 \\ z_4 \end{pmatrix}.
\end{equation}
Let $s_k = s_{k,I}+is_{k,Q}$ for \mbox{$k=1,2,3,4$}. The weight matrices \mbox{$A_{k,I},A_{k,Q}$} of the real symbols \mbox{$s_{k,I}$,$s_{k,Q}$}, upto a sign change, are
\begin{equation*} 
\begin{array}{ll}
A_{1,I} = I_2,  &  A_{1,Q} = iZ, \\
A_{2,I} = ZX,  &  A_{2,Q} = iX, \\
A_{3,I} = Z,  &  A_{3,Q} = iI, \\
A_{4,I} = X,  &  A_{4,Q} = iZX.  
\end{array}
\end{equation*}
This code uses all the $8$ elements of $\Lambda_2$ as weight matrices. From~\eqref{HTWPGA2}, we see that the encoding groups are: \mbox{$\{s_{1,I},s_{1,Q}\}$},\mbox{$\{s_{2,I},s_{2,Q}\}$} and \mbox{$\{s_{3,I},s_{3,Q},s_{4,I},s_{4,Q}\}$}. Since the combined encoding of \mbox{$\{s_{3,I},s_{3,Q},s_{4,I},s_{4,Q}\}$} does not affect the fast-decodability offered by the design, the decoding complexity of the code is $2M^3$ for arbitrary complex constellations and $4M^2$ for complex constellations. However, it was already shown in Section~\ref{NewClass} that the decoding complexity can be further reduced to $3M^2$ when using square QAM.

\section{Discussion} \label{FutureWorkSection}
The approach of this paper is inspired and similar to that of \cite{CRSS}. In this paper we provided a framework for constructing low decoding complexity STBCs from codes over $\mathbb{F}_4$. New FD and FGD codes were provided based on this approach. However, only square designs for $2^m$ number of antennas is considered. Thus the designs obtained from \mbox{$\mathbb{F}_2 \oplus \mathbb{F}_4^m$} for arbitrary number of antennas, by deletion of select columns, will not be delay optimal. In \cite{KaR2} delay optimal codes are provided also for the case when number of transmit antennas is not a power of $2$. These codes can not be obtained from \mbox{$\mathbb{F}_2 \oplus \mathbb{F}_4^m$}. Also the Doubling construction given in~\cite{KiR} for multigroup decodable codes is general and it was applied in that work to STBCs from Division Algebras~\cite{SSS}. In this paper we only obtain its application to codes with Pauli weight matrices. Further the relationship between the presented class of $g$-group decodable designs and those obtainable from Extended Clifford Algebras remains to be explored. 




\section*{Appendix A}

\begin{center}
P\textsc{roof of }T\textsc{heorem}~\ref{InductionStep}
\end{center}


Define a map 
\begin{equation*}
\rho_{n-1} : \mathbb{Z}_{Q_1} \times \dots \times \mathbb{Z}_{Q_n} \to \mathbb{Z}_{Q_1} \times \dots \times \mathbb{Z}_{Q_{n-1}}
\end{equation*}
such that for any $u \in \mathbb{Z}_{Q_1} \times \dots \times \mathbb{Z}_{Q_n}$
\begin{equation*}
\rho_{n-1}((u_1,u_2,\dots,u_n)) = (u_1,u_2,\dots,u_{n-1})
\end{equation*}

The proof is by induction on $Q_{n+1}$. First we show that the theorem is true for $Q_{n+1} = 1$. Then the induction step is proved.

Let $Q_{n+1}=1$. Choose $\mathcal{A}_{n+1} = \{a_{n+1}[0]\}$ where $a_{n+1}[0]$ is any element of $\mathbb{R}$. Let \mbox{$u,v \in \mathbb{Z}_{Q_1} \times \dots \times \mathbb{Z}_{Q_n} \times \mathbb{Z}_1$} and $u \neq v$. Since there is only one element in $\mathbb{Z}_1$ we have that \mbox{$\rho_n(u) \neq \rho_n(v)$}. Using this along with the hypothesis that \mbox{$\mathcal{C}({\bf X_n},\mathcal{A}_1 \times \dots \times \mathcal{A}_n)$} offers full-diversity we have \mbox{$det(C_{n+1}[u] - C_{n+1}[v])$}
\begin{align}
&= det(\sum_{i=1}^{n+1}{(a_i[u_i]A_i - a_i[v_i]A_i)}) \nonumber \\
&= det(\sum_{i=1}^{n}{(a_i[u_i]A_i - a_i[v_i]A_i)}) \nonumber \\
&= det(C_n[\rho_n(u)]-C_n[\rho_n(v)]) \nonumber \\
&\neq 0 \nonumber 
\end{align}

In order to prove the induction step, we assume that the theorem is true for $Q_{n+1}=k$ with the real constellation $\mathcal{A}'_{n+1}$ for some positive integer $k$. We prove that the theorem is true for $Q_{n+1}=k+1$ by appending another point $a_{n+1}[k] \in \mathbb{R}$ to $\mathcal{A}'_{n+1}$. Thus $a_{n+1}[k]$ must not be an element of $\mathcal{A}'_{n+1}$. In order to guarantee full diversity it must satisfy an additional criterion which is, for any \mbox{$u,v \in \mathbb{Z}_{Q_1} \times \dots \times \mathbb{Z}_{Q_n} \times \mathbb{Z}_{k+1}$} and $u \neq v$, \mbox{$det(C_{n+1}[u] - C_{n+1}[v]) \neq 0$}. There are four cases given below. For each of these cases this criterion translates into some condition on $a_{n+1}[k]$. The point to be chosen must satisfy all these criteria and must not be an element of $\mathcal{A}'_{n+1}$.
\begin{enumerate}

\item {\em $u_{n+1} \neq k$ and $v_{n+1} \neq k$ :} In this case 
\begin{equation*}
C_{n+1}[u],C_{n+1}[v] \in \mathcal{C}({\bf X_{n+1}},\mathcal{A}_1 \times \dots \times \mathcal{A}'_{n+1})
\end{equation*}
Since \mbox{$\mathcal{C}({\bf X_{n+1}},\mathcal{A}_1 \times \dots \times \mathcal{A}_{n+1})$} offers full-diversity this case does not impose any condition on $a_{n+1}[k]$.

\item {\em $u_{n+1} = v_{n+1} = k$ :} Together with $u \neq v$ we have \mbox{$\rho_n(u) \neq \rho_n(v)$}. Thus \mbox{$det(C_{n+1}[u] - C_{n+1}[v])$}
\begin{align}
&= det(\sum_{i=1}^{n+1}{(a_i[u_i]A_i - a_i[v_i]A_i)}) \nonumber \\
&= det(\sum_{i=1}^{n}{(a_i[u_i]A_i - a_i[v_i]A_i)}) \nonumber \\
&= det(C_n[\rho_n(u)]-C_n[\rho_n(v)]) \nonumber \\
&\neq 0 \nonumber 
\end{align}
Even this case does not impose any condition on $a_{n+1}[k]$.

\item {\em $u_{n+1} \neq k$ and $v_{n+1} = k$ :} In this case $a_{n+1}[k] \in \mathbb{R}$ must not be a solution of the polynomial equation
	\begin{equation}
	h_{u,v}(z) = det(C_{n+1}[u] - \sum_{i=1}^{n}{a_i[v_i]A_i} - zA_{n+1}) = 0
	\end{equation}
The above polynomial equation is not identically zero i.e., \mbox{$h_{u,v}(z) \in \mathbb{C}[z] \setminus \{0\}$}. This can be shown by considering two cases
	\begin{enumerate}
	
	\item When $\rho_n(u) \neq \rho_n(v)$, we have $h_{u,v}(a_{n+1}[u_{n+1}])$
		\begin{align}
		&= det(C_{n+1}[u] - \sum_{i=1}^{n}{a_i[v_i]A_i} - a_{n+1}[u_{n+1}]A_{n+1}) \nonumber \\
		&= det(C_{n}[\rho_n(u)] - C_{n}[\rho_n(v)]) \nonumber \\
		&\neq 0 \nonumber 
		\end{align}
		
		\item When $\rho_n(u) = \rho_n(v)$, we have $h_{u,v}(z)$
		\begin{align}
		&= det(C_{n+1}[u] - \sum_{i=1}^{n}{a_i[u_i]A_i} - zA_{n+1}) \nonumber \\
		&= det(a_{n+1}[u_{n+1}]A_{n+1}- zA_{n+1}) \nonumber \\
		&= (a_{n+1}[u_{n+1}]-z)^Ndet(A_{n+1}) \nonumber \\
		& \in \mathbb{C}[z] \setminus \{0\} \nonumber 
		\end{align}
				
	\end{enumerate}
\item {\em $u_{n+1} = k$ and $v_{n+1} \neq k$ :} In this case $a_{n+1}[k] \in \mathbb{R}$ must not be a solution of the polynomial equation
	\begin{equation}
	g_{u,v}(z) = det(C_{n+1}[v] - \sum_{i=1}^{n}{a_i[u_i]A_i} - zA_{n+1}) = 0
	\end{equation}
The above polynomial equation is not identically zero i.e., \mbox{$g_{u,v}(z) \in \mathbb{C}[z] \setminus \{0\}$}. The proof of this is similar to the proof in last case.	

\end{enumerate}

Thus $\mathcal{C}({\bf X_{n+1}},\mathcal{A}_1 \times \dots \times \mathcal{A}_{n+1})$ will offer full-diversity if $a_{n+1}[k]$ satisfies all of the following conditions
\begin{enumerate}

\item $a_{n+1}[k] \notin \mathcal{A}'_{n+1}$

\item $a_{n+1}[k]$ is not a root of $h_{u,v}(z)$ for any $u,v$ from case~$3$.

\item $a_{n+1}[k]$ is not a root of $g_{u,v}(z)$ for any $u,v$ from case~$4$. 

\end{enumerate}

Any non-zero polynomial $f(z) \in \mathbb{C}[z]$ has only finitely many solutions in $\mathbb{C}$ and hence only finitely many solutions in $\mathbb{R}$. There are only finitely many such non-zero equations in the above criteria. Also there are only finite number of elements in $\mathcal{A}'_{n+1}$. Thus there are infinitely many choices of $a_{n+1}[k]$ that can make \mbox{$\mathcal{C}({\bf X_{n+1}},\mathcal{A}_1 \times \dots \times \mathcal{A}_{n+1})$} offer full-diversity. This proves the existence of full-diversity, single real symbol encodable code \mbox{$\mathcal{C}({\bf X_{n+1}},\mathcal{A}_1 \times \dots \times \mathcal{A}_{n+1})$} for \mbox{$Q_{n+1}=k+1$}. Thus the induction step is proved. $\blacksquare$


\end{document}